\documentclass[journal, a4paper]{IEEEtran}

\usepackage{algorithm,algorithmic}
\usepackage{amsmath,amssymb,mathrsfs}
\usepackage{array}
\usepackage{booktabs}
\usepackage{cite}
\usepackage{epsfig}
\usepackage{fancyhdr}
\usepackage{float}
\usepackage{graphicx,color}
\usepackage[hidelinks]{hyperref}
\usepackage{mathtools}
\usepackage{multirow}
\usepackage{psfrag}
\usepackage{stfloats}
\usepackage{subfigure}
\usepackage{tikz,siunitx}
\usetikzlibrary{shapes,arrows}
\usepackage{url}
\usepackage{svg}
\usepackage{balance}

\newtheorem{lem}{Lemma}
\newtheorem{thm}{Theorem}

\newtheorem{prop}{Property}

\newtheorem{rmk}{Remark}

\definecolor{sblue}{RGB}{0,51,120}
\definecolor{sred}{RGB}{200,51,130}

\newcommand{\mat}[1]{\mbox{\boldmath $#1$}}

\newcommand{\secref}[1]{Section \ref{#1}}
\renewcommand{\eqref}[1]{(\ref{#1})}

\ifCLASSINFOpdf
\else
\fi

\begin{document}
\title{Accelerating Iteratively Linear Detectors in Multi-User (ELAA-)MIMO Systems with UW-SVD}
\author{Jiuyu Liu, Yi Ma, Jinfei Wang, and Rahim Tafazolli
	\thanks{Jiuyu Liu, Yi Ma (\textit{corresponding author}), Jinfei Wang, and Rahim Tafazolli are with the 5GIC and 6GIC, Institute for Communication Systems (ICS), University of Surrey, Guildford, United Kingdom, GU2 7XH, e-mails: (jiuyu.liu, y.ma, jinfei.wang, r.tafazolli)@surrey.ac.uk.}
	\thanks{This work was partially supported by the UK Department for Science, Innovation and Technology under the Future Open Networks Research Challenge project TUDOR (Towards Ubiquitous 3D Open Resilient Network).}
	\thanks{This work has been partially presented in SPAWC'2023, Shanghai \cite{Liu2023a}.}}
	
\markboth{IEEE Transactions on Wireless Communications}%
{Accelerating Iteratively Linear Detectors in Multi-User (ELAA-)MIMO Systems with UW-SVD}
\maketitle

\begin{abstract}
Current iterative multiple-input multiple-output (MIMO) detectors suffer from slow convergence when the wireless channel is ill-conditioned.
The ill-conditioning is mainly caused by spatial correlation between channel columns corresponding to the same user equipment, known as intra-user interference.
In addition, in the emerging MIMO systems using an extremely large aperture array (ELAA), spatial non-stationarity can make the channel even more ill-conditioned.
In this paper, user-wise singular value decomposition (UW-SVD) is proposed to accelerate the convergence of iterative MIMO detectors.
Its basic principle is to perform SVD on each user's sub-channel matrix to eliminate intra-user interference.
Then, the MIMO signal model is effectively transformed into an equivalent signal (e-signal) model, comprising an e-channel matrix and an e-signal vector.
Existing iterative algorithms can be used to recover the e-signal vector, which undergoes post-processing to obtain the signal vector.
It is proven that the e-channel matrix is better conditioned than the original MIMO channel for spatially correlated (ELAA-)MIMO channels.
This implies that UW-SVD can accelerate current iterative algorithms, which is confirmed by our simulation results.
Specifically, it can speed up convergence by up to $10$ times in both uncoded and coded systems.
\end{abstract}
                                

\section{Introduction} \label{sec01}
\IEEEPARstart{T}{he} primary focus of this paper is low-complexity signal detection for multi-user multiple-input multiple-output (MIMO) systems, particularly those deployed with extremely-large aperture arrays (ELAA).
ELAA-MIMO systems can increase spectral efficiency by more than tenfold over current massive-MIMO systems \cite{Bjoernson2017}.
This is because users are typically located in the near-field of the ELAA; and the near-field channels can provide higher spatial resolution compared to the far-field massive-MIMO channels \cite{Ji2023, Cui2023, Ouyang2023}.
For instance, under strong line-of-sight (LoS) conditions, ELAA-MIMO can support multiple data streams from a user equipment (UE) equipped with multiple antennas, while massive-MIMO channels can only support a single data stream per UE \cite{Wu2023, Dardari2020}.
In massive-MIMO systems, the wireless channel can become ill-conditioned due to high spatial correlation among channel columns \cite{Loyka2001}.
However, ELAA channels can be even more ill-conditioned due to both channel spatial correlation and non-stationarity \cite{Elzanaty2023}.
This makes the design of low-complexity MIMO detectors challenging, particularly those iterative algorithms with square-order complexity \cite{BJORNSON20193}.

Maximum-likelihood (ML) detector, while achieving the optimal detection performance, is computationally impractical due to its exponentially growing complexity \cite{Chang2017}.
Linear detectors, such as zero-forcing (ZF) and minimum mean square error (LMMSE), offer a more computationally efficient alternative, providing near-optimal detection performance \cite{Wang2022a}.
However, they both require a Gram matrix inverse with cubic-order complexity, which limits their applications in large-scale MIMO systems \cite{Wang2023b}.
Instead, iterative algorithms achieve ZF/LMMSE detection performance with square-order complexity, bypassing the matrix inverse.
Conventional algorithms with simple structures, such as Richardson iteration (RI) \cite{Gao2014} and Neumann series \cite{Zhu2015}, can offer fast convergence in well-conditioned channel matrices.
Conversely, they may exhibit divergence when applied to ill-conditioned channel matrices \cite{Wang2023b}.
This challenge motivates more advanced algorithms that aim to achieve fast convergence in such channel matrices.

\subsection{Relevant Prior Arts} \label{sec01a}
Current iterative algorithms can be classified into three main categories \cite{Yang2015, Albreem2019, Axelsson1985}: \textit{1)} gradient methods, \textit{2)} belief propagation, and \textit{3)} matrix-splitting (MS) based methods.

Gradient methods can achieve global convergence in solving the problem of linear MIMO detection \cite{Li2022}.
For instances, steepest descent (SD) method updates in the same direction as RI, but converges faster than RI because it optimizes the step size in each iteration \cite{Qin2016};
conjugate gradient (CG) method leverages the Hermitian nature of the Gram channel matrix to determine a more efficient update direction, which further accelerates convergence compared to SD \cite{Yin2014, Liu2020}.
In addition, quasi-Newton (QN) methods represent an important branch of gradient methods, such as symmetric rank 1 (SR1) \cite{Wang2023c} and Broyden-Fletcher-Goldfarb-Shanno (BFGS) \cite{Li2022}.
Due to the iterative approximation of the Gram matrix inversion, QN methods typically exhibit cubic order complexity.
Recently, the application of limited-memory BFGS (L-BFGS) to linear MIMO detection has demonstrated its ability to achieve convergence equivalent to that of BFGS while requiring only square-order complexity \cite{Li2022a}.

Belief propagation refers to iterative message passing (MP) algorithms.
Among these algorithms, approximate MP (AMP) was initially proposed for compressive sensing and has been applied for MIMO detection in recent years \cite{Donoho2009, Lyu2019}.
The complexity of AMP is comparable to that of L-BFGS, and both algorithms exhibit similar convergence rate in massive-MIMO systems.
However, AMP diverges in ELAA-MIMO systems due to the channel non-stationarity \cite{Liu2023e}.
AMP variants such as orthogonal AMP (OAMP) and vector AMP (VAMP) \cite{Ma2017, Rangan2019} have been proposed to address this problem, and their detection performance is even slightly better than that of LMMSE \cite{He2018}.
However, they both introduce computational overhead due to the requirement for matrix inverse or singular value decomposition (SVD), resulting in cubic-order complexity.

In MS-based methods, the Gram channel matrix is divided into the sum of several individually invertible sub-matrices, the inversions of which will be used to accelerate the convergence \cite{Albreem2019}.
Typically, they divide the Gram matrix into its diagonal part and its upper and lower triangular parts.
MS-based methods include Jacobi iteration (JI) \cite{Kong2016}, Gauss-Seidel (GS) method \cite{Zhang2021}, and successive over-relaxation (SSOR) \cite{Xie2016}.
Specifically, the inverses of the diagonal and lower triangular matrices are used for the JI and GS methods, respectively.
Furthermore, SSOR converges faster than the JI and GS methods because it uses the inverses of both the upper and lower triangular matrices to further accelerate the convergence.
The triangular matrix inverse has square-order complexity, making it scalable for large-scale MIMO systems \cite{Albreem2019}.

\subsection{Motivation of This Paper} \label{sec01b}
Based on recent theoretical advancements \cite{Nam2017, Liu2021, Liu2024a, Zhang2023, Lu2023} and empirical field measurements \cite{Martinez2014, Harris2016, Yuan2022, Karstensen2022}, it has been observed that intra-user interference is much stronger than inter-user interference in ELAA-MIMO systems.
This phenomenon also holds true for conventional massive-MIMO systems when spatial correlations are taken into account (see \cite{Loyka2001} and our discussion in Section \ref{sec04b} for more details).
However, current iterative algorithms typically use a generalized approach to tackle the problem of channel ill-conditioning, disregarding the distinctive features of multi-user MIMO channels.
As a result, when spatial correlation is considered, current algorithms still require tens of iterations to converge in (ELAA-)MIMO systems \cite{Liu2023a}.
This motivates the rest of this paper.

\subsection{Contributions of This Paper} \label{sec01c}
\begin{figure}[t]
	\subfigure{
	\begin{minipage}[t]{0.48\textwidth}
		\centering
		\includegraphics[width=0.95\textwidth]{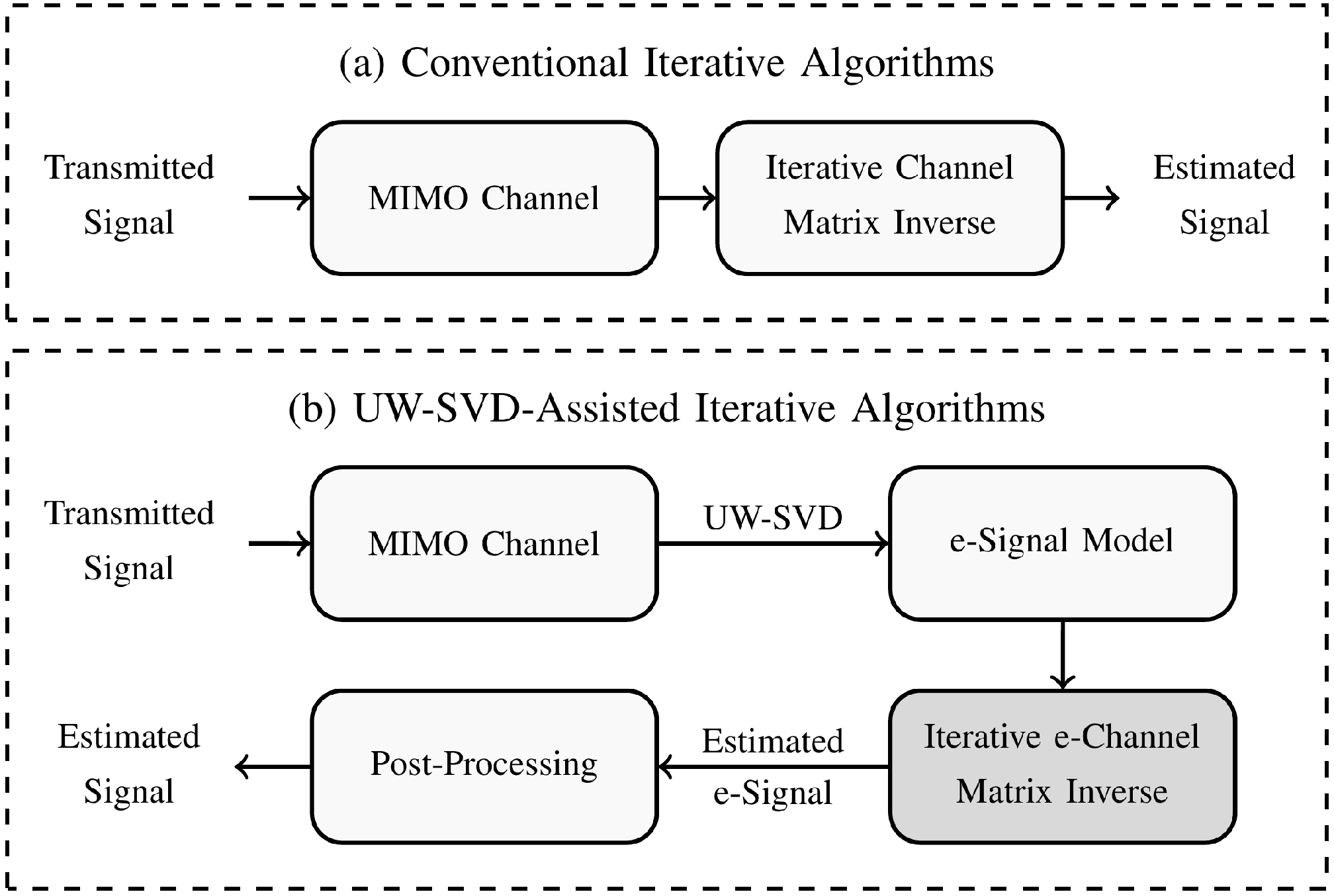}
	\end{minipage}}
	\caption{\label{fig01}Illustration of the major differences between the current iterative algorithms and UW-SVD-assisted iterative algorithms.}
	\vspace{-1em}
\end{figure}

In this paper, we propose to utilize user-wise SVD (UW-SVD) to accelerate the convergence of current iterative algorithms in multi-user (ELAA-)MIMO systems.
The concept of UW-SVD is to perform SVD on the sub-channel matrix corresponding to each UE \footnote{\label{footnote1} The authors are aware that this option is used in some prior works, e.g., \cite{Lou2013,Li2017,Khamidullina2022}, but they all focus on optimizing power allocation strategies using the singular value matrices.
In contrast, our study focuses on the left unitary matrices to accelerate the convergence of current iterative algorithms.}, thereby eliminating the intra-user interference.
The MIMO signal model can then be transformed to an equivalent signal (e-signal) model containing an e-channel matrix and a corresponding e-signal vector.
The major differences between current iterative algorithms and UW-SVD-assisted iterative algorithms are illustrated in Fig. \ref{fig01}.
It can be observed that current algorithms applied to the MIMO signal model converge directly to the estimation of the transmitted signal.
In contrast, UW-SVD-based algorithms first converge to an estimation of the e-signal vector and then convert it back to the transmitted signal through a post-processing step.
An e-signal model-based ZF detector, termed e-ZF, was developed in our previous work \cite{Liu2023a}.
It is proven that, after post-processing, e-ZF detector can provide equivalent estimation to ZF detector.

In addition to \cite{Liu2023a}, an LMMSE detector for the e-signal model, termed e-LMMSE, is developed in this paper.
Also, it is proven to provide equivalent detection performance to the LMMSE detector.
Furthermore, it is demonstrated that the e-channel matrix exhibits a lower condition number compared to the original channel matrix, particularly in ELAA-MIMO systems.
Considering an ELAA-MIMO system under LoS conditions as an example, when the spatial correlation of small-scale fading is not accounted for, the condition number of the MIMO channel matrix is approximately $60$, while the condition number of the e-channel matrix is significantly lower at approximately $5$. 
Moreover, when the spatial correlation is considered, the condition number of the MIMO channel matrix increases substantially to approximately $700$. 
However, even in this scenario, the condition number of the e-channel matrix remains significantly lower at approximately $7$.
A lower condition number indicates that a matrix is less sensitive to perturbations, which means that iterative algorithms can converge to the correct solution more quickly.
Therefore, the proposed UW-SVD can significantly accelerate the convergence of current iterative algorithms to achieve ZF/LMMSE performance. 
This is evident in our computer simulations. 
For example, the UW-SVD-assisted SSOR converges ten times faster than SSOR in both uncoded and coded ELAA-MIMO systems.
Finally, it is worth noting that UW-SVD can also speed up the convergence in conventional massive-MIMO channels when the spatial correlation is taken into account.

\subsection{Organization and Notations}
The rest of this paper is organized as follows.
Section \ref{sec02} presents the system model, preliminaries, and problem statement.
Section \ref{sec03} describes the principle of UW-SVD and its application in accelerating the convergence of current iterative algorithms.
Section \ref{sec04} presents the convergence analysis.
Section \ref{sec05} presents the numerical and simulation results.
Finally, the conclusion is presented in Section \ref{sec06}.
\subsubsection*{Notations}
Regular letter, lower-case bold letter, and capital bold letter represent scalar, vector and matrix, respectively.
The notations $[\cdot]^{H}$, $[\cdot]^{-1}$, $\|\cdot\|$, $\mathbb{E}\{\cdot\}$ and $\mathrm{cond}(\cdot)$, represent the Hermitian, inverse, Euclidean norm, expectation, and condition number of a matrix (a vector or a scalar if appropriate), respectively.
$\mathbb{D}(\cdot)$ and $\mathbb{L}(\cdot)$ denote a matrix formed by the diagonal and lower-triangular part of a matrix, respectively.
$\lambda_{\text{max}}(\cdot)$ or $\lambda_{\text{min}}(\cdot)$ denote the maximum or minimum eigenvalue of a matrix.
$\mathrm{diag}(\cdot)$ constructs the input matrices in a block diagonal form.
$\mathbf{I}$ and $\mathbf{0}$ denote identity and zero matrices with compatible dimensions.

\section{System Model, Preliminaries, and Problem Statement} \label{sec02}
This section begins by introducing the system model.
Next, it presents linear MIMO detectors and low-complexity iterative algorithms. 
Finally, it discusses the challenges of these algorithms in achieving ZF/LMMSE detection performance in ill-conditioned channel matrix.

\subsection{System Model}
Let $M$ and $N$ denote the number of service antennas and user antennas, respectively.
For ELAA-MIMO and massive-MIMO systems, their signal models share the same mathematical form and can be expressed as follows
\begin{equation} \label{eqn01}
	\mathbf{y} = \mathbf{H}\mathbf{x} + \mathbf{z},
\end{equation}
where $\mathbf{y} \in \mathbb{C}^{M \times 1}$ denotes the received signal vector, $\mathbf{x} \in \mathbb{C}^{N \times 1}$ the transmitted signal vector, $\mathbf{z} \sim \mathcal{CN}(0,\sigma_z^2\mathbf{I})$ the additive white Gaussian noise (AWGN), $\sigma_{z}^{2}$ the noise variance, and $\mathbf{I}$ represents an identity matrix with compatible dimensions.
Each element of $\mathbf{x}$ is drawn from a finite alphabet-set with equal probability and fulfills: $\mathbb{E}\{\mathbf{x}\} = \mathbf{0}$ and $\mathbb{E}\{\mathbf{x}\mathbf{x}^{H}\} = \sigma_{x}^{2}\mathbf{I}$.
Note that the random channel matrix $\mathbf{H} \in \mathbb{C}^{M \times N}$ has different distributions in massive-MIMO and ELAA-MIMO systems.
In Section \ref{sec05a}, we consider four random distributions of $\mathbf{ H}$ for computer simulations.

In ELAA-MIMO and conventional massive-MIMO systems, the performance can be significantly degraded by spatial correlation between user antennas.
This spatial correlation leads to two types of interference: \textit{1)} intra-user interference, and \textit{2)} inter-user interference.
Intra-user interference occurs when signals transmitted from different antenna elements to the same user interfere with each other due to spatial correlation.
Conversely, inter-user interference is caused by signals intended for other users.
Typically, intra-user interference is much stronger than inter-user interference.
The reason for this is that the distance between antennas serving the same user is usually less than the distance between antennas serving different users.
Section \ref{sec04b} provides a mathematical justification for this phenomenon.
Consequently, the primary objective of this work is to mitigate the intra-user interference for both ELAA-MIMO and massive-MIMO systems.

\subsection{Preliminaries}
The two most classical linear MIMO detectors are ZF and LMMSE, which can be expressed as follows \cite{Albreem2019}
\begin{equation} \label{eqn19410911}
	\widehat{\mathbf{x}} = \mathbf{A}^{-1} \mathbf{b},
\end{equation}
where $\mathbf{b} = \mathbf{H}^{H}\mathbf{y}$ represents the matched filter vector, $\widehat{\mathbf{x}}$ the estimation of $\mathbf{x}$, and $\mathbf{A}$ is a Gram filter matrix.
For ZF and LMMSE detectors, $\mathbf{A}$ can be expressed as follows
\begin{equation} \label{eqn10531708}
	\mathbf{A}=\left\{\begin{array}{l}
		\mathbf{A}_{\textsc{zf}} \triangleq  \mathbf{H}^{H}\mathbf{H};  \\
		\mathbf{A}_{\textsc{lmmse}} \triangleq  \mathbf{H}^{H}\mathbf{H} + \rho^{-1} \mathbf{I},
	\end{array}\right.
\end{equation}
where $\rho = \sigma_{x}^{2} / \sigma_{z}^{2}$ denotes the signal-to-noise ratio (SNR).
However, both ZF and LMMSE detectors require the inverse calculation of $\mathbf{A}$, which is computationally prohibitive for large MIMO sizes.

A number of iterative algorithms have been proposed to efficiently solve the problem in \eqref{eqn19410911} bypassing $\mathbf{A}$ inverse \cite{Yang2015, Albreem2019, Axelsson1985}.
Their general form can be expressed as follows
\begin{equation} \label{eqn10361211}
	\mathbf{x}_{t+1} = f(\mathbf{x}_{t}; \mathbf{A}, \mathbf{b}),
\end{equation}
where $\mathbf{x}_{t}$ represents the $t^{th}$ estimation of $\mathbf{x}$ and $f(\cdot)$ is a linear function that varies depending on the specific algorithm employed.
Taking RI as an example, $f_{\textsc{ri}}(\cdot)$ is given by \cite{Gao2014}
\begin{equation} \label{eqn14372611}
	f_{\textsc{ri}}(\mathbf{x}_{t}; \mathbf{A}, \mathbf{b}) = \mathbf{x}_{t} + (\mathbf{b} - \mathbf{A}\mathbf{x}_{t}),
\end{equation}
which has a simple structure, but it diverges when $\mathbf{A}$ is ill-conditioned \cite{Wang2022a}.

To address this issue, more advanced algorithms have been proposed to achieve faster convergence.
For instance, the iterative process of MS-based methods is given by \cite{Liu2023a}
\begin{equation} \label{eqn11241510}
	f_{\textsc{ms}}(\mathbf{x}_{t}; \mathbf{A}, \mathbf{b}) = \mathbf{x}_{t} + \mathbf{M}^{-1}(\mathbf{b} - \mathbf{A}\mathbf{x}_{t}),
\end{equation}
where $\mathbf{M}$ represents the preconditioning matrix, and it is constructed based on the following matrix splitting
\begin{equation} \label{eqn19562011}
	\mathbf{A} = \mathbb{L}(\mathbf{A}) + \mathbb{L}(\mathbf{A})^{H} -  \mathbb{D}(\mathbf{A}),
\end{equation}
where $\mathbb{L}(\cdot)$ and $\mathbb{D}(\cdot)$ represent matrices formed by the lower triangular and diagonal parts of the input matrix, respectively.
Since $\mathbf{A}_{\textsc{zf}}$ and $\mathbf{A}_{\textsc{lmmse}}$ are both Hermitian matrices, $\mathbb{L}(\mathbf{A})^{H}$ in \eqref{eqn19562011} actually represents the upper triangular part of $\mathbf{A}$.
For JI, GS and SSOR methods, the preconditioning matrices are defined as follows: $\mathbf{M}_{\textsc{ji}} = \mathbb{D}(\mathbf{A})$, $\mathbf{M}_{\textsc{gs}} = \mathbb{L}(\mathbf{A})$ \cite{Zhang2021}, and $\mathbf{M}_{\textsc{ssor}} = \mathbb{L}(\mathbf{A}) \mathbb{D}(\mathbf{A})^{-1}\mathbb{L}(\mathbf{A})^{H}$ \cite{Xie2016}, respectively.
MS-based methods generally have faster convergence than RI due to the use of $\mathbf{M}$, which enables a more efficient update direction.
In addition, the inversion of triangular matrices exhibits square-order complexity, making it computationally efficient.

Furthermore, gradient methods can also provide accelerated convergence over RI by using adaptive optimization of the step size, update direction, or both.
To avoid redundancy within this paper, a detailed discussion of gradient methods will be presented in Section \ref{sec03d}.

\subsection{Problem Statement}
The convergence rate of the iterative algorithms described in \eqref{eqn10361211} is significantly influenced by the condition number of $\mathbf{A}$ \cite{Albreem2019}.
Specifically, for a given iterative function, it converges faster when the condition number is smaller.
In this paper, the condition number is defined as follows
\begin{equation}
	\mathrm{cond}(\mathbf{A}) \triangleq \dfrac{\lambda_{\text{max}}(\mathbf{A})}{\lambda_{\text{min}}(\mathbf{A})},
\end{equation} 
where $\lambda_{\text{max}}(\cdot)$ and $\lambda_{\text{min}}(\cdot)$ represent the maximum and minimum eigenvalues of the input matrix, respectively.
In massive-MIMO systems without spatial correlation, current algorithms can offer fast convergence, since $\mathbf{A}$ is well-conditioned.
However, they all demonstrate slow convergence in ELAA-MIMO systems, particularly in scenarios dominated by LoS links \cite{Liu2023a}.
The reason is that ELAA channel matrices could be very ill-conditioned, meaning $\mathrm{cond}(\mathbf{A}) \gg 1$ \cite{Liu2024}.
As discussed in Section \ref{sec01b}, the main reason contributing to (ELAA-)MIMO channel ill-conditioning is the strong intra-user interference.
Therefore, the objective of this paper is to efficiently eliminate the impact of intra-user interference on the iterative process, and the following sections are motivated.

\section{UW-SVD-Assisted Iterative Algorithms} \label{sec03}
This section introduces the concept of UW-SVD and its role in transforming the MIMO signal model into an e-signal model. 
Subsequently, the derivations of e-ZF and e-LMMSE detectors for the e-signal model are presented. 
Furthermore, existing iterative algorithms are employed to estimate the e-signal vector, which is then transformed back to the estimation of $\mathbf{x}$ through the post-processing step.

\subsection{The Concept of UW-SVD}
Suppose there are $K$ UEs deployed in the MIMO system, and the $k^{th}$ UE is equipped with $N_{k}$ antennas.
The system configuration satisfies: $\sum_{k=1}^{K} N_k = N$.
Then, the complete channel matrix can be represented in a concatenated format as follows
\begin{equation} \label{eqn21381310}
	\mathbf{H} = [\mathbf{H}_{1}, ... , \mathbf{H}_{K}],
\end{equation}
where $\mathbf{H}_{k} \in \mathbb{C}^{M\times N_k}$ represents the sub-channel matrix corresponding to the $k^{th}$ UE.
To eliminate intra-user interference, we apply the economy-size SVD \footnote{A variant of SVD that computes only the necessary components for tall matrices, enhancing computational efficiency \cite{Shaik2023}.} to each user's sub-channel matrix as follows
\begin{equation} \label{eqn21371310}
	\mathbf{H}_{k} = \mathbf{U}_{k}\mathbf{\Sigma}_{k}\mathbf{V}^{H}_{k},
\end{equation}
where $\mathbf{U}_k \in \mathbb{C}^{M \times N_k}$ represents the left unitary matrix, $\mathbf{\Sigma}_k \in \mathbb{R}^{N_k \times N_k}$ the diagonal matrix containing the singular values, and $\mathbf{V}_k \in \mathbb{C}^{N_k \times N_k}$ represents the right unitary matrix.
This step is the so-called UW-SVD \textsuperscript{\ref{footnote1}}.

Substituting \eqref{eqn21371310} into \eqref{eqn21381310} with some tidy-up work, $\mathbf{H}$ can be decomposed into the following three matrix multiplications
\begin{equation} \label{eqn09470308}
	\mathbf{H} = \mathbf{\Psi} \mathbf{\Sigma} \mathbf{V}^{H},
\end{equation}
where $\mathbf{\Psi} \triangleq [\mathbf{U}_{1}, ..., \mathbf{U}_{K}]$ represents the concatenation of $\mathbf{U}_{k}$. $\mathbf{\Sigma} \triangleq \mathrm{diag}(\mathbf{\Sigma}_{1}, \dots, \mathbf{\Sigma}_{K})$ and $\mathbf{V} \triangleq \mathrm{diag}(\mathbf{V}_{1}, \dots, \mathbf{V}_{K})$ are block diagonal matrices containing the singular value and right-unitary matrices, respectively.
The notation $\mathrm{diag}(\cdot)$ represents the construction of the input matrices in a block diagonal manner.
It is obvious that $\mathbf{\Sigma}$ is a positive-real diagonal matrix and $\mathbf{V}$ is a unitary matrix, i.e.,
\begin{equation}
	\mathbf{V}^{H}\mathbf{V} = \mathbf{V}\mathbf{V}^{H} = \mathbf{I}.
\end{equation}
However, it is worth noting that $\mathbf{\Psi}$ is not a unitary matrix in practical MIMO systems.
This is because the left-unitary matrices are tall matrices, and $\mathbf{U}_{k}$ for different UEs may not necessarily be orthogonal to each other.
Next, we will explore the transformation from the MIMO signal model to e-signal model using UW-SVD.

\subsection{The e-Signal Model}
According to the UW-SVD in \eqref{eqn09470308}, the MIMO signal model in \eqref{eqn01} can be transformed into an e-signal model as follows
\begin{equation}\label{eqn16542104}
	\mathbf{y} = \mathbf{\Psi} \mathbf{s} + \mathbf{z}, 
\end{equation}
where $\mathbf{s} \triangleq \mathbf{\Sigma}\mathbf{V}^{H} \mathbf{x}$ represents the e-signal vector, and $\mathbf{\Psi}$ represents the e-channel matrix.
Therefore, $\mathbf{\Psi}$ and $\mathbf{s}$ are the linear representations of $\mathbf{H}$ and $\mathbf{x}$, respectively.

\begin{rmk}
	With this e-signal model, it is important to understand the properties of $\mathbf{s}$ and $\mathbf{\Psi}$.
	Taking $\mathbf{s}$ as an example, its expectation and covariance can be expressed as follows
	\begin{IEEEeqnarray}{rl}
		\mathbb{E}\{\mathbf{s}\}\ &= \mathbf{0}; \label{eqn14342511} \\
		\mathbb{E}\{\mathbf{s}\mathbf{s}^{H}\}\ &= \sigma_{x}^{2} \mathbf{\Sigma}^{2}, \label{eqn16301811}
	\end{IEEEeqnarray}
	which can be easily obtained from $\mathbb{E}\{\mathbf{x}\} = \mathbf{0}$ and $\mathbb{E}\{\mathbf{x}\mathbf{x}^{H}\} = \sigma_{x}^{2} \mathbf{I}$.
	This indicates that distinct e-signal data streams are orthogonal to each other and they exhibit different transmission powers.
	Moreover, we have the following
	\begin{equation} \label{eqn09242002}
		\mathbb{D}\big(\mathbf{\Psi}^{H}\mathbf{\Psi}\big) = \mathbf{I},
	\end{equation}
	which indicates that the complexity of computing $\mathbf{D}\big(\mathbf{\Psi}^{H}\mathbf{\Psi}\big)$ can be ignored in certain iterative algorithms, such as JI and L-BFGS methods.
\end{rmk}

Note that the condition number of $\mathbf{\Psi}$ is crucial for this paper because it significantly affects the convergence of the UW-SVD-assisted algorithms.
This property will be examined in Section \ref{sec04}, where we present a comprehensive convergence analysis.
Before that, we focus on the development of linear detectors for the e-signal model in the next subsection.

\subsection{The e-ZF and e-LMMSE Detectors} \label{sec03c}
In this subsection, we develop the e-ZF and e-LMMSE detectors for the e-signal model.
Additionally, it is demonstrated that they can achieve the same detection performance as the corresponding ZF or LMMSE detector after a low-complexity post-processing step.

Given that the e-signal model is a linear representation of the MIMO signal model, its two linear detectors (i.e., e-ZF and e-LMMSE) can be expressed in the following general form
\begin{equation} \label{eqn15221111}
	\widehat{\mathbf{s}} = \mathbf{\Phi}^{-1} \mat{\delta},
\end{equation}
where $\mat{\delta} = \mathbf{\Psi}^{H}\mathbf{y}$ denotes the matched filter vector for the e-signal model.
Similar to that in (\ref{eqn10531708}), $\mathbf{\Phi}$ for e-ZF and e-LMMSE can be expressed as follows
\begin{equation} \label{eqn17011811}
	\mathbf{\Phi}=\left\{\begin{array}{l}
		\mathbf{\Phi}_{\textsc{zf}} \triangleq \mathbf{\Psi}^{H}\mathbf{\Psi};  \\
		\mathbf{\Phi}_{\textsc{lmmse}} \triangleq  \mathbf{\Psi}^{H}\mathbf{\Psi} + \rho^{-1} \mathbf{\Sigma}^{-2}.
	\end{array}\right.
\end{equation}
It is obvious that \eqref{eqn15221111} and \eqref{eqn19410911} share the same mathematical structure.
Hence, any iterative algorithm designed to determine $\widehat{\mathbf{x}}_{\textsc{zf}}$ or $\widehat{\mathbf{x}}_{\textsc{lmmse}}$ can be directly applied to determine $\widehat{\mathbf{s}}_{\textsc{zf}}$ or $\widehat{\mathbf{s}}_{\textsc{lmmse}}$, respectively.
Since the objective of MIMO signal detection is to reconstruct the transmitted signal vector $\mathbf{x}$, a post-processing step is required to convert $\widehat{\mathbf{s}}$ back to $\widehat{\mathbf{x}}$.

\textit{Post-Processing Step:}
Consistent with the definition of $\mathbf{s}$, we propose to reconvert $\widehat{\mathbf{x}}$ from $\widehat{\mathbf{s}}$ as follows
\begin{equation} \label{eqn15121011}
	\widehat{\mathbf{x}} = \mathbf{V}\mathbf{\Sigma}^{-1} \widehat{\mathbf{s}}.
\end{equation}
where $\mathbf{\Sigma}$ is a diagonal matrix, so that computing $\mathbf{\Sigma}^{-1}$ requires only linear computational complexity.
Plugging (\ref{eqn17011811}) into (\ref{eqn15121011}) with some tidy-up works, we can have the following
\begin{IEEEeqnarray}{c}
	\widehat{\mathbf{x}}_{\textsc{zf}} = \mathbf{V}\mathbf{\Sigma}^{-1}\widehat{\mathbf{s}}_{\textsc{zf}}; \label{eqn21151811} \\
	\widehat{\mathbf{x}}_{\textsc{lmmse}} = \mathbf{V}\mathbf{\Sigma}^{-1}\widehat{\mathbf{s}}_{\textsc{lmmse}}. \label{eqn21051811}
\end{IEEEeqnarray}
With the post-processing step, the e-ZF and e-LMMSE detectors can provide the same detection performance as the ZF and LMMSE detectors, respectively.
This implies that any iterative algorithm that converges to $\widehat{\mathbf{s}}$ can provide ZF or LMMSE detection performance.
The specific steps of the UW-SVD-assisted algorithms are discussed in the next section.

\subsection{UW-SVD Assisted Iterative Algorithms} \label{sec03d}
As discussed in Section \ref{sec03c}, all the MS-based methods and gradient methods can be directly applied to estimate $\mathbf{s}$ \footnote{AMP and its variants require further modifications to determine $\mathbf{s}$ due to their structures; this exploration is beyond the scope of this paper.
Additionally, using MS-based methods and L-BFGS is sufficient to demonstrate the advantage of the proposed UW-SVD method.}.
These methods share the same iterative structure as \eqref{eqn10361211}, except that the specific parameters are adjusted as follows
\begin{equation}
	\mathbf{s}_{t+1} = f(\mathbf{s}_{t}; \mathbf{\Phi}, \mat{\delta}),
\end{equation}
which is the so-called UW-SVD-assisted iterative algorithms.
In the case where $\mathbf{\Phi} = \mathbf{\Phi}_{\textsc{zf}}$, $\mathbf{s}_{t}$ will converge to the e-ZF solution. 
Conversely, when $\mathbf{\Phi} = \mathbf{\Phi}_{\textsc{lmmse}}$, $\mathbf{s}_{t}$ will converge to the e-LMMSE solution.
It is worth noting that the convergence rate of UW-SVD-assisted algorithms is dominated by $\mathrm{cond}(\mathbf{\Phi})$, rather than $\mathrm{cond}(\mathbf{A})$.
The comparison between $\mathrm{cond}(\mathbf{\Phi})$ and $\mathrm{cond}(\mathbf{A})$ will be comprehensively explored in Section \ref{sec05}.

Similar to \eqref{eqn11241510}, MS-based methods assisted by UW-SVD can be expressed as follows
\begin{equation} \label{eqn10251911}
	f_{\textsc{ms}}(\mathbf{s}_{t}; \mathbf{\Phi}, \mat{\delta}) = \mathbf{s}_{t} + \mathbf{M}^{-1}(\mat{\delta} - \mathbf{\Phi}\mathbf{s}_{t}),
\end{equation}
where the preconditioning matrix $\mathbf{M}$ should be constructed based on the matrix splitting of $\mathbf{\Phi}$.
For the case $\mathbf{\Phi} = \mathbf{\Phi}_{\textsc{zf}}$, JI and SSOR methods can be further simplified.
Specifically, JI and RI are equivalent to each other because $\mathbf{M}_{\textsc{ji}} = \mathbb{D}(\mathbf{\Phi}_{\textsc{zf}})$ is an identity matrix; and $\mathbf{M}$ for SSOR can be simplified to $\mathbf{M}_{\textsc{ssor}} = \mathbb{L}(\mathbf{\Phi}_{\textsc{zf}}) \mathbb{L}(\mathbf{\Phi}_{\textsc{zf}})^{H}$.

Gradient methods can also be employed to address the problem in \eqref{eqn15221111}, such as SD, CG and L-BFGS.
It is demonstrated that L-BFGS converges faster than SD while maintaining similar square-order complexity \cite{Li2022a}.
Moreover, it is proven that L-BFGS and CG are equivalent when solving the convex MIMO detection problem \cite{Nazareth1979}.
Therefore, we consider L-BFGS as an example of gradient methods in this paper. 
Its iterative process is given by \cite{Li2022}
\begin{equation} \label{eqn14511510}
	f_{\textsc{lbfgs}}(\mathbf{s}_{t}; \mathbf{\Phi}, \mat{\delta}) = \mathbf{s}_{t} + \xi_{t} \mathbf{d}_{t},
\end{equation}
where $\xi_{t}$ denotes the step size as follows
\begin{equation} \label{eqn08302202}
	\xi_{t} = - \dfrac{\mathbf{g}_{t}^{H} \mathbf{d}_{t}}{\mathbf{d}_{t}^{H}\mathbf{\Phi}\mathbf{d}_{t}},
\end{equation}
and $\mathbf{d}_{t}$ denotes the update direction as follows
\begin{equation}
	\mathbf{d}_{t} = \mathbf{\Theta}_{t}\mathbf{g}_{t},
\end{equation}
where $\mathbf{g}_{t} \triangleq \mathbf{\Phi} \mathbf{s}_{t} - \mat{\delta}$ denotes the gradient direction. 
$\mathbf{\Theta}_{t}$ represents the approximation of Hessian matrix, and it can be expressed as follows
\begin{equation} \label{eqn11561510}
	\mathbf{\Theta}_{t}= \bigg(\dfrac{(\mathbf{s}_{t} - \mathbf{s}_{t-1}) (\mathbf{g}_{t} - \mathbf{g}_{t-1})^{H}}{(\mathbf{s}_{t} - \mathbf{s}_{t-1})^{H} (\mathbf{g}_{t} - \mathbf{g}_{t-1})} - \mathbf{I} \bigg) \mathbf{\Theta}_{0},
\end{equation}
where $\mathbf{\Theta}_{0}$ represents the initial approximation.
Typically, $\mathbf{\Theta}_{0}$ is set as $\mathbb{D}(\mathbf{\Phi})^{-1}$.
For e-ZF detector, the term $\mathbf{\Theta}_{0}$ in \eqref{eqn11561510} can be omitted, since $\mathbb{D}(\mathbf{\Phi}_{\textsc{zf}})^{-1} = \mathbf{I}$.

\begin{algorithm}[t]
	{{\renewcommand{\thealgorithm}{}
	\small
	\caption{UW-SVD assisted L-BFGS algorithm}
	\begin{algorithmic}[1]\label{alg01}
		\renewcommand{\algorithmicrequire}{\textbf{Input:}} 
		\REQUIRE~\\ $\mathbf{y}$: received signal vector;
		\\ $\mathbf{H}$: MIMO channel matrix;
		\\ $\rho$: SNR;
		\\ $T$: number of iterations;
		\\ $\mathbf{s}_{0} = \mathbf{0}$: the initialization vector.
	    	\renewcommand{\algorithmicrequire}{\textbf{Output:}} 
		\REQUIRE~\\$\widehat{\mathbf{x}}$: the estimation of $\mathbf{x}$;
		\renewcommand{\algorithmicensure}{\textbf{START}}
		\ENSURE  
		\STATE {\bf let} $t =0$; call \eqref{eqn09470308} to compute $\mathbf{\Psi}$, $\mathbf{\Sigma}$, and $\mathbf{V}$;
		\STATE {\bf let} $\mat{\delta} = \mathbf{\Psi}^{H}\mathbf{y}$;  {\bf let} $\mathbf{\Phi} = \mathbf{\Psi}^{H}\mathbf{\Psi} + \rho^{-1} \mathbf{\Sigma}^{-2}$;
		\STATE call \eqref{eqn14511510} to compute $\mathbf{s}_{t+1}$; {\bf then} $t \leftarrow t+1$;
		\STATE {\bf repeat} step 3 until $t = T$; 
		\STATE call \eqref{eqn15121011} to compute $\widehat{\mathbf{x}}$;
		\renewcommand{\algorithmicensure}{\textbf{END}}
		\ENSURE
	\end{algorithmic}}}
\end{algorithm}

\textit{Pseudocode: }
The UW-SVD assisted L-BFGS algorithm is presented in the \textbf{Algorithm}.
It can provide the LMMSE detection performance, since the filter matrix $\mathbf{\Phi}$ is set to be $\mathbf{\Phi}_{\textsc{lmmse}}$.
By setting $\mathbf{\Phi} = \mathbf{\Psi}^{H}\mathbf{\Psi}$ in step $2$, the algorithm will provide ZF detection performance.
In step $3$, \eqref{eqn14511510} is the iterative function of L-BFGS.
Therefore, if it is replaced by \eqref{eqn10251911}, the \textbf{Algorithm} would become UW-SVD assisted MS-based methods.
Step $5$ is the post-processing step.
It aims to reconvert $\widehat{\mathbf{x}}$ from $\widehat{\mathbf{s}}$.
Additionally, UW-SVD can accelerate the convergence of numerous other iterative algorithms, such as SD and CG.
The proposed UW-SVD method leverages sample structure to facilitate their application in accelerating the convergence of various existing algorithms.

\subsection{Complexity Analysis}
\begin{table*}
	\centering
	\caption{Complexity Analysis of UW-SVD and Various Iterative MIMO Detectors}
	\label{tab01}
	\renewcommand{\arraystretch}{1.2}
	\resizebox{0.7\textwidth}{!}{
	\begin{tabular}{@{}ccccc@{}}
		\toprule
		Algorithms & Calculation of $\mathbf{A}$ or $\mathbf{\Phi}$ & Matrix Inverse & Per Iteration    & UW-SVD   \\ \midrule
		ZF/LMMSE   & $MN^2$                                           & $N^3$          & 0                & \multirow{6}{*}{$N_{\textsc{ue}} MN + N_{\textsc{ue}}N + 2N$} \\
		RI         & $0$                                              & $0$            & $2MN$            &                                                               \\
		JI         & $0$                                              & $0$            & $2MN + N$        &                                                               \\
		GS         & $MN^2$                                           & $N^2$          & $1.5 N^2$ &                                                               \\
		SSOR       & $MN^2$                                           & $N^2$          & $2N^2 + N$  &                                                               \\
		L-BFGS     & $0$                                              & $0$            & $4MN + N^2 + 5N$ &                                                               \\
		\bottomrule
	\end{tabular}%
	}
\vspace{-0em}	
\end{table*}

The objective of this section is to demonstrate that the proposed UW-SVD method has low computational-complexity.
To simplify and clarify the complexity analysis, we adopt a common assumption in multi-user MIMO systems with $K$ users, where each user has $N_{\textsc{ue}}$ antennas, i.e., $KN_{\textsc{ue}} = N$.
The computational complexity of UW-SVD assisted iterative algorithms can be divided into three main parts: UW-SVD, post-processing, and iterative process.
We start the complexity analysis from the UW-SVD and post-processing steps.

Performing SVD on $\mathbf{H}_{k}$ has a complexity of $MN_{\textsc{ue}}^{2}$ \cite{Golub1996}, resulting in a total complexity of $KMN_{\textsc{ue}}^{2}$ for all the users.
Also, the complexity of UW-SVD can also be expressed as $N_{\textsc{ue}}MN$, since $KN_{\textsc{ue}} = N$.
In the post-processing step, computing $\mathbf{\Sigma}^{-1}$ has a complexity of $N$, while the computation of $[\mathbf{\Sigma}^{-1}\mathbf{s}]$ has the same complexity of $N$.
Moreover, given that $\mathbf{V}$ is a block diagonal matrix, the complexity of calculating $\mathbf{V}[\mathbf{\Sigma}^{-1}\mathbf{s}_{t}]$ is $KN_{\textsc{ue}}^{2} = N_{\textsc{ue}}N$.
Therefore, the overall complexity of the post-processing step is $N_{\textsc{ue}}N + 2N$.
Furthermore, the total complexity of UW-SVD together with the post-processing step is $N_{\textsc{ue}}MN + N_{\textsc{ue}}N + 2N$.
In MIMO systems, the number of antennas per UE is usually small, typically $N_{\textsc{ue}} = 2$ or $4$.
Hence, the complexity of UW-SVD method stays at the quadratic order.

It is worth noting that not all the iterative algorithms require the computation of $\mathbf{A}$ or $\mathbf{\Phi}$, including RI, JI, and L-BFGS methods.
Taking RI as an example, if we replace $\mathbf{\Phi}_{\textsc{zf}} = \mathbf{\Psi}^{H}\mathbf{\Psi}$ in (\ref{eqn10251911}), its iterative process can be expressed as follows
\begin{equation} \label{eqn09012002}
	f_{\textsc{ri}}(\mathbf{s}_{t}; \mathbf{\Phi}, \mat{\delta}) = \mathbf{s}_{t} + (\mat{\delta} - \mathbf{\Psi}^{H}\mathbf{\Psi} \mathbf{s}_{t}),
\end{equation}
where we can first compute $[\mathbf{\Psi}\mathbf{s}_{t}]$ with a complexity of $MN$, and then compute $\mathbf{\Psi}^{H}[\mathbf{\Psi}\mathbf{s}_{t}]$ with another complexity of $MN$.
In this successive manner, the calculation of $\mathbf{\Phi}$ can be avoided.
This can also be applied to all the other iterative methods, such as the calculation of $\xi_{t}$ in (\ref{eqn08302202}) in L-BFGS method.
In addition, a similar complexity can be obtained by replacing $\mathbf{\Phi}_{\textsc{zf}}$ with $\mathbf{\Phi}_{\textsc{lmmse}}$ in (\ref{eqn09012002}).
The complexity of computing $\mathbf{\Sigma}^{-1}$ is only $N$, since it is a diagonal matrix.
Furthermore, the calculation of $\mathbb{D}(\mathbf{\Psi}^{H}\mathbf{\Psi})^{-1}$ in JI and L-BFGS methods can be ignored because it is an identity matrix according to (\ref{eqn09242002}).

The authors are aware that certain iterative algorithms require the calculation of $\mathbf{A}$ or $\mathbf{\Phi}$, such as the GS and SSOR methods.
This is because they need to compute $\mathbb{L}(\mathbf{A})^{-1}$ or $\mathbb{L}(\mathbf{\Phi})^{-1}$.
Furthermore, the complexity of computing $\mathbb{L}(\mathbf{A})^{-1}$ or $\mathbb{L}(\mathbf{\Phi})^{-1}$ is $N^2$ due to the triangular structure.
The complexity of these iterative algorithms, in short, remains essentially the same whether UW-SVD is applied or not.
Moreover, ZF and LMMSE detectors require the computation of $\mathbf{A}$ with a cubic-order complexity of $MN^2$.
They also necessitate the calculation of $\mathbf{A}^{-1}$ with cubic-order complexity of $N^3$.
TABLE \ref{tab01} summarizes the complexity of UW-SVD and various MIMO detectors.
The matrix inverse operation is the primary reason why ZF/LMMSE is impractical for real-time signal processing due to its serially computational complexity of cubic order \cite{Albreem2019}.
This motivates traditional iterative methods to (partially) circumvent the need for matrix inversion.

\textbf{Discussion of complexity reduction:}
Our simulation results demonstrate that the proposed UW-SVD method can reduce the computational complexity of SSOR and L-BFGS methods by up to $90\%$ and $67\%$, respectively (see Figs. \ref{fig05d} and \ref{fig07}).
This substantial reduction in complexity is primarily achieved by decreasing the number of iterations required, despite the additional complexity introduced by UW-SVD.
As shown in TABLE \ref{tab01}, the extra computational burden imposed by UW-SVD is equivalent to $16$ SSOR iterations or a single L-BFGS iteration.
However, the advantages of UW-SVD far outweigh its cost.
For instance, the result in Fig. \ref{fig05d} shows that UW-SVD can accelerate SSOR by up to approximately $240$ iterations.
Moreover, our simulation results in Fig. \ref{fig07} demonstrate that UW-SVD can accelerate L-BFGS method by up to $13$ iterations.
This significant acceleration in convergence of UW-SVD more than offsets its additional processing cost, thus significantly reducing the overall computational complexity.

\section{Convergence Analysis} \label{sec04}
In this section, the objective is to compare $\mathrm{cond}(\mathbf{A})$ and $\mathrm{cond}(\mathbf{\Phi})$ in both massive-MIMO and ELAA-MIMO systems.
The next two subsections provide detailed results of the comparison in each system, respectively.

\subsection{Massive-MIMO with i.i.d. Rayleigh Fading Channels} \label{sec03a}
To better understand the relationship between $\mathrm{cond}(\mathbf{A})$ and $\mathrm{cond}(\mathbf{\Phi})$, we first introduce the following concept of favorable propagation in massive-MIMO systems.

\begin{prop} [Favorable Propagation \cite{Ngo2014}] \label{lem01}
Suppose that elements of $\mathbf{H}$ to follow independent and identically distributed (i.i.d.) Rayleigh fading as (\ref{eqn13502111}), given $N_{k}, \forall k$, as $M$ tends to infinity, we have the following
\begin{equation}\label{eqn21441211}
	\lim\limits_{M \rightarrow \infty} \mathbf{H}_{k}^{H}\mathbf{H}_{k} = \mathbf{I}, \quad \forall k. 
\end{equation}
\end{prop}

\begin{thm}\label{thm03}
Suppose that every element of $\mathbf{H}$ obeys an i.i.d. Rayleigh distribution in \eqref{eqn13502111}, given $N_{k}, \forall k$, as $M$ tends to infinity, we have the following
\begin{IEEEeqnarray}{ll}
& \lim\limits_{M \rightarrow \infty} \mathrm{cond}(\mathbf{\Phi}_{\textsc{zf}}) =  \mathrm{cond}(\mathbf{A}_{\textsc{zf}}); \label{eqn22142711a} \\
& \lim\limits_{M \rightarrow \infty} \mathrm{cond}(\mathbf{\Phi}_{\textsc{lmmse}}) =  \mathrm{cond}(\mathbf{A}_{\textsc{lmmse}}). \label{eqn22142711b}
\end{IEEEeqnarray}
\end{thm}

\begin{IEEEproof}
See \textsc{Appendix} \ref{appThm03}
\end{IEEEproof}

\textit{Theorem} \ref{thm03} implies that the UW-SVD-assisted algorithm has a comparable convergence rate comparable to the existing algorithm.
The reason is that the intra-user interference in i.i.d. Rayleigh fading channels is very weak, which could limit the gain of UW-SVD.
This theoretical finding is verified in the numerical results of \textbf{Experiment 1} in Section \ref{sec05c}.
On the contrary, intra-user interference is strong in spatially correlated (ELAA-)MIMO systems, especially in the presence of LoS links.
In the next subsection, we will show that $\mathrm{cond}(\mathbf{\Phi}) < \mathrm{cond}(\mathbf{A})$ in such channels.

\subsection{Spatially Correlated (ELAA-)MIMO Channels} \label{sec04b}
Given that UW-SVD aims to address intra-user interference, our focus lies primarily on understanding the user-side spatial correlation \footnote{\label{foot4}In this section, we focus on the user-side spatial correlation to facilitate the theoretical analysis. However, in our simulations, we adopt the Kronecker model in (\ref{eqn20572802}), which considers both user-side and BS-side spatial correlations.}, which is defined as follows \cite{Loyka2001}
\begin{equation}
	\mathbf{R}_{\textsc{ue}} \triangleq \mathbb{E}\{\mathbf{H}^{H}\mathbf{H}\},
\end{equation}
where $\mathbf{R}_{\textsc{ue}}$ is usually described by an exponential correlation matrix in conventional massive-MIMO systems.
For example, if two user antennas are situated at a distance of $d$, the correlation between these two antennas can be expressed as follows \cite{Loyka2001}
\begin{equation} \label{eqn13280203}
	r(d) = \exp (-d / \mu),
\end{equation}
where $\mu \geq 0$ represents the scaling factor.
In multi-user MIMO systems, the distance between different users is typically much greater than the distance between antennas belonging to the same UE.
Therefore, we have the following assumption:
\begin{equation}\label{eqn17032102}
	\textit{A1):}\quad	\mathbf{R}_{\textsc{ue}}^{k, j}  = \mathbf{0}, \quad \forall k \neq j,
\end{equation}
where $\mathbf{R}_{\textsc{ue}}^{k, j} \in \mathbb{R}^{N_{k} \times N_{j}}$ denotes a block of $\mathbf{R}_{\textsc{ue}}$ representing the correlation between user $k$ and user $j$.

Let us take an example to validate this assumption.
Suppose we have a UE equipped with two antennas spaced apart by half the carrier wavelength. 
Assuming the carrier frequency is $3.5$ $\mathrm{GHz}$ and the parameter $\mu$ equals $0.2$, the correlation between the two intra-user antennas is $r(0.0429) \approx 0.8$. 
This suggests a significant spatial correlation between the two intra-user antennas.
In contrast, when considering two distinct users separated by one meter, their correlation $r(1) \approx 4.5 \times 10^{-5}$ implies nearly orthogonal behavior.
In real-world scenarios, user distances typically exceed one meter. 
Therefore, the assumption \textit{A1)} in (\ref{eqn17032102}) is validated for practical MIMO systems.

\begin{lem} \label{lem02}
	Given \textit{A1}, suppose the MIMO channel is $\mathbf{H} = \mathbf{\Omega} \sqrt{\mathbf{R}_\textsc{ue}}$, where each element of $\mathbf{\Omega}$ follows an i.i.d. Rayleigh distribution, we have the following
	\begin{equation}\label{eqn12222102}
		\lim\limits_{M \rightarrow \infty} \mathbf{H}_{k}^{H} \mathbf{H}_{j} = \mathbf{0}, \quad \forall k \neq j,
	\end{equation}
	where $\mathbf{R}_{\textsc{ue}}^{k, k} \in \mathbb{R}^{N_{k} \times N_{k}}$ is a block of $\mathbf{R}_{\textsc{ue}}$ representing the correlation between the antenna elements of user $k$.
\end{lem}

\begin{IEEEproof}
	See \textsc{Appendix} \ref{appLem02}.
\end{IEEEproof}

In massive-MIMO systems, the number of service-antennas can be in the hundreds or even thousands.
Consequently, the condition in (\ref{eqn12222102}) can be approximated as follows
\begin{equation}
	\textit{A2):} \quad \mathbf{H}_{k}^{H} \mathbf{H}_{j} = \mathbf{0}, \quad \forall k \neq j
\end{equation}
Moreover, as discussed in section \ref{sec01b}, in ELAA-MIMO systems, intra-user interference is much stronger than inter-user interference.
This is because the dominant power of different users can be received by different service antennas.
This phenomenon is known as spatial orthogonality \cite{Sanguinetti2020, Wu2023}.
Therefore, it can be assumed that \textit{A2} also holds in ELAA-MIMO systems.
With assumption \textit{A2}, our focus shifts to the comparison between $\mathrm{cond}(\mathbf{A})$ and $\mathrm{cond}(\mathbf{\Phi})$ for ZF and LMMSE detectors, respectively.

\begin{thm} \label{thm04}
Given \textit{A2}, it can be obtained that the condition number of $\mathbf{A}_{\textsc{zf}}$ is larger than that of $\mathbf{\Phi}_{\textsc{zf}}$, i.e.,
	\begin{equation} \label{eqn10352811} 
	\mathrm{cond}(\mathbf{\Phi}_{\textsc{zf}}) < \mathrm{cond}(\mathbf{A}_{\textsc{zf}}).
	\end{equation}
\end{thm}

\begin{IEEEproof}
	See \textsc{Appendix} \ref{appThm04}
\end{IEEEproof}

\begin{thm} \label{thm05}
Given \textit{A2}, suppose that the transmitted power is normalized to $1$, i.e., $\sigma_{x}^{2} = 1$, we have the following
	\begin{equation} \label{eqn11042811}
		\mathrm{cond}(\mathbf{\Phi}_{\textsc{lmmse}}) < \mathrm{cond}(\mathbf{A}_{\textsc{lmmse}}),
	\end{equation}
	when 
	\begin{equation}\label{eqn20201911}
		\sqrt{\mathrm{cond}(\mathbf{A}_{\textsc{zf}})} \lambda_{\text{min}}(\mathbf{A}_{\textsc{zf}})  > \sigma_{z}^{2}.
	\end{equation}
\end{thm}

\begin{IEEEproof}
	See \textsc{Appendix} \ref{appThm05}
\end{IEEEproof}

\begin{rmk} \label{rmk02}
	\textit{Theorem} \ref{thm04} implies that UW-SVD-assisted algorithms can provide faster convergence to ZF detection performance compared to current algorithms.
	\textit{Theorem} \ref{thm05} suggests the similar conclusion for LMMSE detectors but with the constraint that the SNR should be greater than a threshold.
	Note that in any MIMO system for multiplexing transmission, it is typically necessary for the minimum eigenvalue of the channel gain to exceed the noise power, i.e., $\lambda_{\text{min}}(\mathbf{A}_{\textsc{zf}}) > \sigma_{z}^{2}$.
	Also, the condition number should be greater than $1$ based on its definition.
	Therefore, the inequality in (\ref{eqn20201911}) will always be satisfied in practical MIMO systems if we aim for acceptable detection performance using multiplexing techniques.
\end{rmk}

\begin{rmk} \label{rmk03}
	\textit{Theorem} \ref{thm04} and \textit{Theorem} \ref{thm05} imply that $M$ is much greater than $N$ or approaches infinity.
	However, deploying such a large number of service antennas is economically impractical in real-world scenarios.
	This necessitates the determination of specific, implementable ranges for $M$ and $N$.
	However, the stochastic nature of (ELAA-)MIMO channels poses a significant challenge. It is impossible to mathematically derive an exact formula for the $M/N$ ratio at which UW-SVD outperforms conventional methods.
	To address this, we turn to experimental results for insights into practical $M/N$ ratios.
	Our experiments, detailed in \secref{sec05}, reveal that UW-SVD achieves significant gains when the $M/N$ ratio is $4$ or $8$.
	These ratios are not only feasible but also commonly found in MIMO systems.
	This alignment between our findings and real-world parameters ensures the applicability of our methods to practical implementations.
\end{rmk}

\section{Numerical and Simulation Results} \label{sec05}
In this section, the objectives are \textit{1)} to compare $\mathrm{cond}(\mathbf{\Phi})$ and $\mathrm{cond}(\mathbf{A})$ in various types of (ELAA-)MIMO channels; \textit{2)} to demonstrate that UW-SVD accelerates the convergence of current algorithms; and \textit{3)} to establish that the advantages observed in uncoded MIMO systems also apply to coded MIMO systems.
This motivates the following three subsections.

\subsection{Channel Models} \label{sec05a}
\textbf{Model 1:} 
In massive-MIMO systems, each element of $\mathbf{H}$ is usually assumed to obey i.i.d. Rayleigh fading as follows
\begin{equation} \label{eqn13502111}
	H_{m,n} = \omega_{m,n} \sim \mathcal{CN}(0,1/M),
\end{equation}
where $1/M$ denotes the normalized variance of each channel element. 
This indicates that the propagation environment is in non-LoS (NLoS) state \footnote{In LoS state, the massive-MIMO channels can also be described by i.i.d. Rician fading.
However, the far-field Rician channel cannot support multiple data streams per UE \cite{Zhang2023, Lu2023}, and are therefore not the scope of this paper.}.

\textbf{Model 2:}
Spherical wavefront should be taken into account for ELAA channel modeling.
In NLoS state, \eqref{eqn13502111} should be extended to i.n.d. (n. for non-identical) Rayleigh fading as follows \cite{Amiri2018}
\begin{equation} \label{eqn11412508}
	H_{m,n} = H_{m,n}^{(0)} \triangleq \Bigg(\dfrac{\beta^{(0)}}{d_{m,n}^{\gamma^{(0)}}} \Bigg) \omega_{m,n},
\end{equation}
where $d_{m,n}$ denotes the distance between the $m^{th}$ service-antenna and the $n^{th}$ user antenna; $\beta^{(0)}$ and $\gamma^{(0)}$ represent the NLoS path-loss coefficient and exponent, respectively.

\textbf{Model 3:} 
Similarly, the ELAA channel in LoS state is described to obey i.n.d. Rician fading as follows \cite{Liu2021}
\begin{equation} \label{eqn14262111}
	H_{m,n} = H_{m,n}^{(1)} \triangleq \dfrac{\beta^{(1)}}{d_{m,n}^{\gamma^{(1)}}}\Bigg(\sqrt{\dfrac{\kappa}{\kappa + 1}} \varphi_{m,n} + \sqrt{\dfrac{1}{\kappa + 1}}\omega_{m,n}\Bigg).
\end{equation}
where $\beta^{(1)}$ and $\gamma^{(1)}$ represent the LoS path-loss coefficient and exponent, respectively, $\kappa$ denotes the Rician K-factor, $\varphi_{m,n} = \exp(-j\frac{2\pi}{\vartheta}d_{m,n})$ the phase of direct LoS link, and $\vartheta$ denotes the wavelength of the carrier wave.

\textbf{Model 4:}
ELAA channel could allow a mixed of LoS and NLoS links due to the large aperture \cite{Liu2024a}.
Each element of $\mathbf{H}$ in this case can be expressed as follows
\begin{equation}
	H_{m,n} = \epsilon_{m,n}^{(\eta_{m,n})} H_{m,n}^{(\eta_{m,n})},
\end{equation}
where $\eta_{m,n} \in \{0, 1\}$ is a binary random variable, with $\eta_{m,n} = 0$ indicates the NLoS state with $H_{m,n}^{(\eta_{m,n})}$ turning into $H_{m,n}^{(0)}$ in \eqref{eqn13502111}, or otherwise $\eta_{m,n} = 1$ indicates the LoS state with $H_{m,n}^{(\eta_{m,n})}$ turning into $H_{m,n}^{(1)}$ in \eqref{eqn14262111}; $\epsilon_{m,n}$ denotes the shadowing effects.
The spatial correlations of LoS/NLoS states and shadowing effects are described by exponentially decaying window \cite{Liu2024a}.
This channel model can yield computer-simulated data that fit well with real-world measurement data, e.g, \cite{Martinez2014, Harris2016}.
Therefore, we employ this ELAA channel model to conduct computer simulations.

\textbf{Kronecker Model:}
Let $\mathbf{\Omega} \in \mathbb{C}^{M \times N}$ be an i.i.d. complex Gaussian matrix, where its $(m, n)$-th element is denoted by $\omega_{m,n}$, as defined in (\ref{eqn13502111}).
The four channel models above can be converted to their spatially correlated versions by replacing $\mathbf{\Omega}$ with $\mathbf{\Omega}_{\text{kron}}$, as follows \cite{Loyka2001}
\begin{equation} \label{eqn20572802}
	\mathbf{\Omega}_{\text{kron}} = \sqrt{\mathbf{R}_{\textsc{bs}}} \mathbf{\Omega} \sqrt{\mathbf{R}_{\textsc{ue}}},
\end{equation}
where $\mathbf{R}_{\textsc{bs}} \in \mathbb{R}^{M \times M}$ and $\mathbf{R}_{\textsc{ue}} \in \mathbb{R}^{N \times N}$ are both exponential correlation matrices representing the BS and UE side correlations, respectively. 	
In MIMO systems, the minimum distance between two antennas should be $\vartheta/2$.
Therefore, we define the following
\begin{equation}
	\varrho \triangleq r(\vartheta/2),
\end{equation}
where $\varrho$ is the spatial correlation between the two closest antennas.
Additionally, when $\varrho = 0$, it means that the small-scale fading of distinct user-to-service antenna links is generated independently.

\subsection{Baselines} \label{sec05b}
The following iterative algorithms are set as baselines for our simulations: GS, SSOR, and L-BFGS.
AMP is not used as a baseline because it converges slower than the L-BFGS method and diverges in the ELAA channel.
Additionally, it must be modified further to recover the e-signal vector.
Due to the page limitations, we cannot demonstrate all the iterative algorithms that proposed in the last sixty years \cite{Yang2015}.
However, the baselines in this section are sufficient to demonstrate the advantages of the proposed UW-SVD method.

\subsection{System Setup and Experiments} \label{sec05c}
The carrier frequency is set to be $3.5$ $\mathrm{GHz}$.
The service array is configured as a uniformly linear array (ULA)\footnote{An exception is Fig. \ref{fig07}, where the service antenna array is configured as a uniform planar array (UPA) with $M = 16 \times 16$ antennas.} with spacing at half the wavelength.
The users are deployed parallel with the ULA at a perpendicular distance of $15$ meters.
Each user is equipped with $N_{\textsc{ue}}$ antennas spaced at half the wavelength.
The maximum distance between two users is set to be $30$ meters.
To ensure a fair comparison of different types of channel models, we normalize the channel gain for each UE, i.e., $\|\mathbf{H}_{k}\|^{2} = N_{\textsc{ue}}, {\forall k}$.
This normalization does not change intra-user interference, which is the primary focus of this paper.
The wireless environment is assumed to be urban-micro street canyon, and the propagation parameters are determined according to the 3rd Generation Partnership Project (3GPP) technical report \cite{3gpp.38.901}, as follows: $\beta^{(0)} = 0.020$, $\gamma^{(0)} = 1.765$, $\beta^{(1)} = 0.007$, $\gamma^{(1)} = 1.050$, $\kappa = 9$ $\mathrm{dB}$ for \textbf{Model 3}, $\kappa \sim \mathcal{LN} (9\ \mathrm{dB}, 10\ \mathrm{dB})$ for \textbf{Model 4}.
The objectives of this section set the following three experiments.

\begin{figure*}[htp]
	\centering
	\subfigure[$\varrho = 0$]{
	\begin{minipage}[t]{0.32\textwidth}
			\centering
			\includegraphics[width=0.95\textwidth]{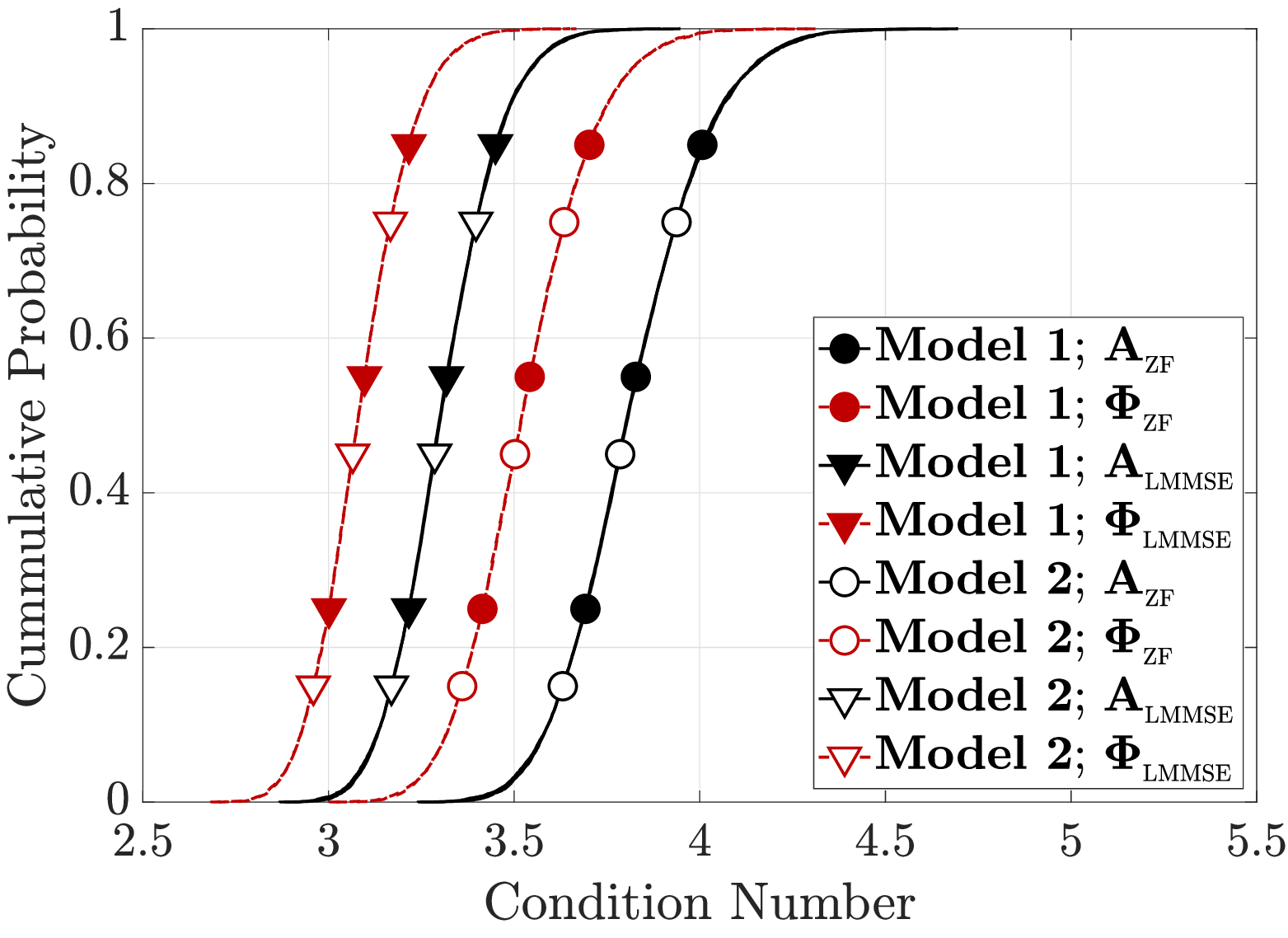}
			\label{fig02a}   	
	\end{minipage}}
	\subfigure[$\varrho = 0.5$]{
		\begin{minipage}[t]{0.32\textwidth}
			\centering
			\includegraphics[width=0.95\textwidth]{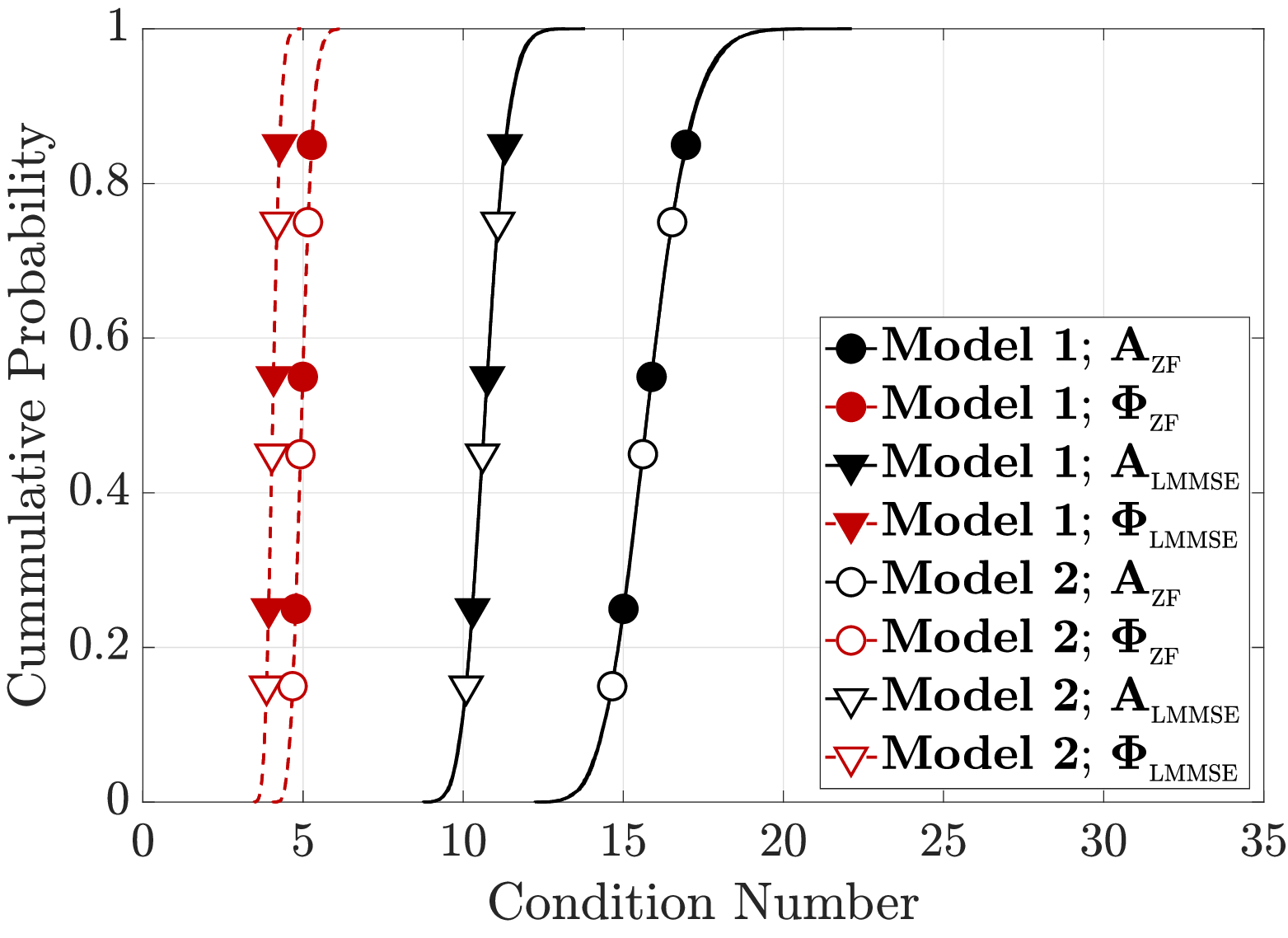}
			\label{fig02b}
	\end{minipage}}	
	\subfigure[$\varrho = 0.8$]{
		\begin{minipage}[t]{0.32\textwidth}
			\centering
			\includegraphics[width=0.95\textwidth]{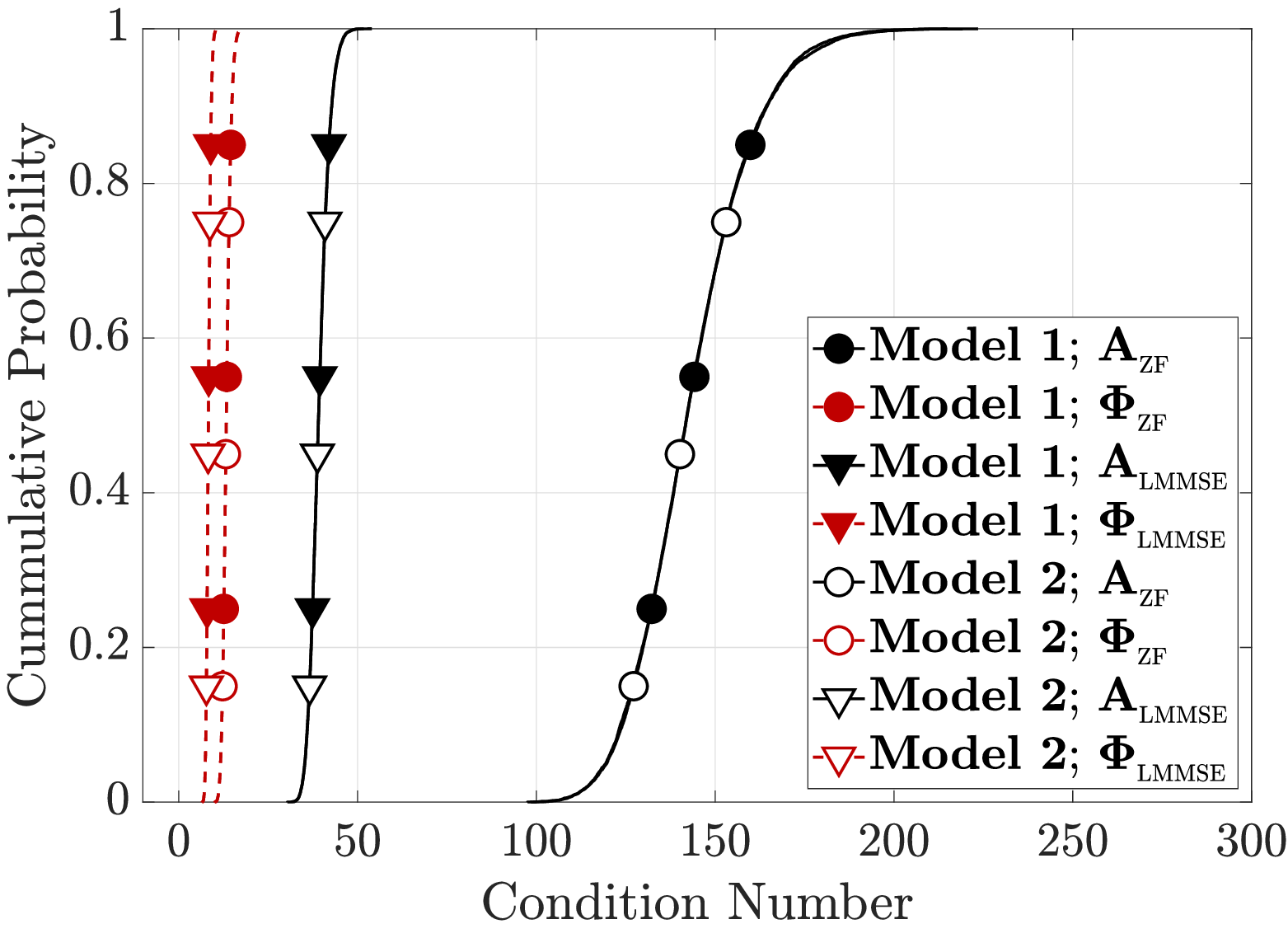}
			\label{fig02c}	   	
	\end{minipage}}
	\caption{\label{fig02}The comparison of $\mathrm{cond}(\mathbf{A})$ and $\mathrm{cond}(\mathbf{\Phi})$ in \textbf{Model 1} and \textbf{Model 2}; $M = 256$; $K = 8$; $N_{\textsc{ue}} = 4$; $\rho = 10$ $\mathrm{dB}$. $\mathrm{cond}(\mathbf{\Phi})$ is smaller than $\mathrm{cond}(\mathbf{A})$ for both ZF/LMMSE detectors, especially in correlated MIMO channels. The matrices in \textbf{Model 1} and \textbf{Model 2} have almost the same condition numbers.}
	\vspace{-0.5em}
\end{figure*}

\begin{figure*}[htp]
	\centering
	\subfigure{
	\begin{minipage}[t]{1\textwidth}
		\centering
		\includegraphics[width=0.95\textwidth]{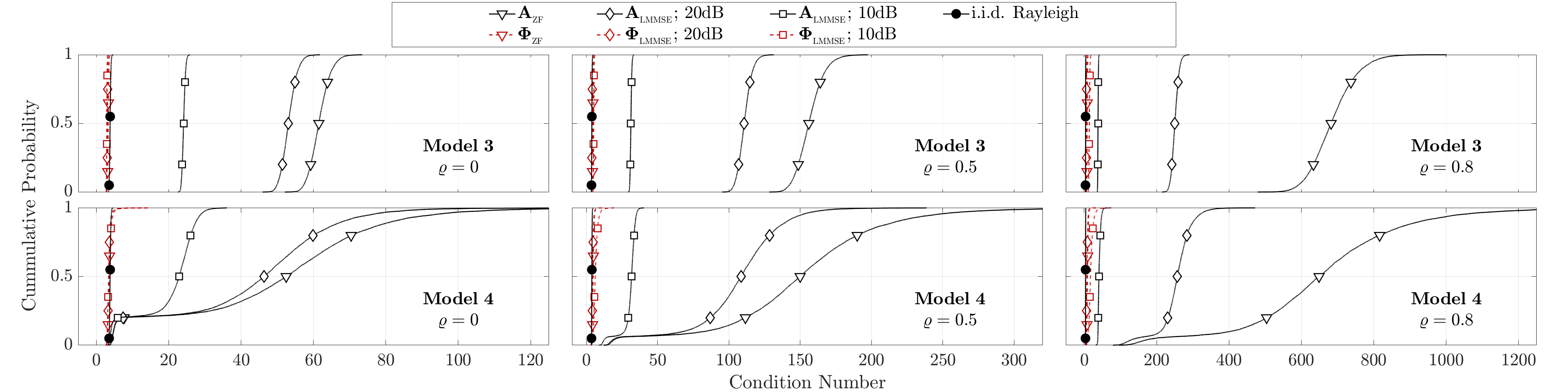}
	\end{minipage}}
	\caption{\label{fig03}The comparison of $\mathrm{cond}(\mathbf{A})$ and $\mathrm{cond}(\mathbf{\Phi})$ in \textbf{Model 3} and \textbf{Model 4}; $M = 256$; $K = 8$; $N_{\textsc{ue}} = 4$. $\mathrm{cond}(\mathbf{\Phi})$ is much smaller than $\mathrm{cond}(\mathbf{A})$ for both ZF/LMMSE detectors in the presence of LoS links. The condition number of $\mathbf{\Phi}$ is similar to that of i.i.d. Rayleigh channel at different correlation levels.}
	\vspace{-1.2em}
\end{figure*}

\textbf{Experiment 1:} 
The objective is to demonstrate that the relationship between $\mathrm{cond}(\mathbf{A})$ and $\mathrm{cond}(\mathbf{\Phi})$ is consistent with the theoretical analysis presented in Section \ref{sec04}.
The cumulative distribution functions (CDFs) of the condition numbers are shown in Figs. \ref{fig02} and \ref{fig03}.
In these two figures, there are $M = 256$ service antennas and $K = 8$ UEs, each equipped with $N_{\textsc{ue}} = 4$ antennas.
In Fig. \ref{fig02a}, where the channel elements are generated independently, it can be seen that $\mathrm{cond}(\mathbf{\Phi})$ is only slightly smaller than $\mathrm{cond}(\mathbf{A})$ in both \textbf{Model 1} and \textbf{Model 2}.
Note that all the condition numbers in this figure are relatively small, which means that numerous iterative algorithms can achieve fast convergence.
However, it is more practical to consider spatial correlations and such results are shown in Figs. \ref{fig02b} and \ref{fig02c}.
By comparing these two figures, it can be observed that $\mathbf{\Phi}$ is better conditioned than $\mathbf{A}$, especially in highly correlated MIMO channels.
This implies that the advantage of the proposed UW-SVD method will be more evident when the correlation increases.

\begin{figure*}[t]
	\centering
	\subfigure[$\varrho = 0$; $\rho = 18$ $\mathrm{dB}$]
	{\begin{minipage}[t]{0.32\textwidth}
			\centering
			\includegraphics[width=0.95\textwidth]{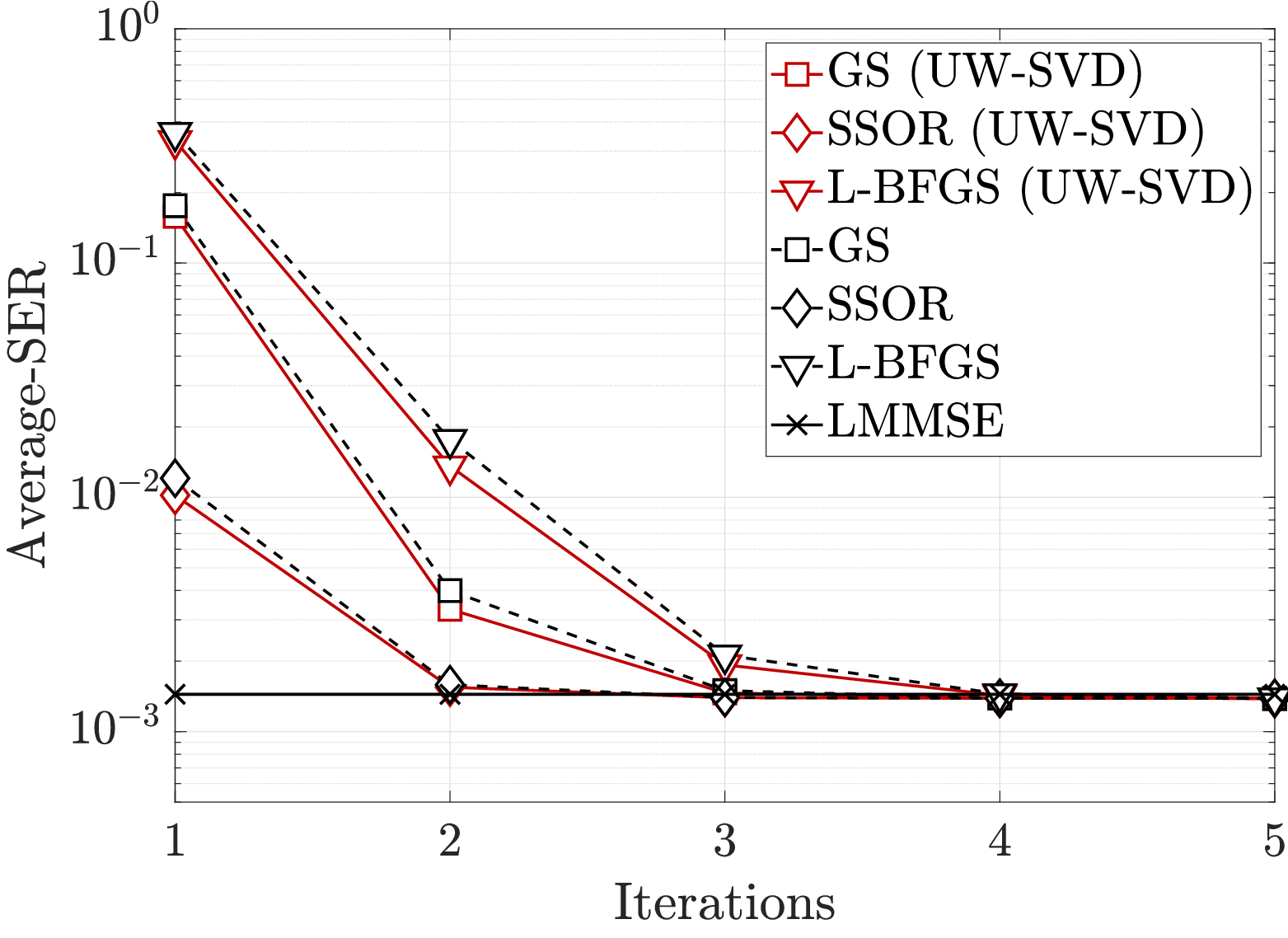}	
			\label{fig04a}   	
	\end{minipage}}		
	\subfigure[$\varrho = 0.5$; $\rho = 20.5$ $\mathrm{dB}$]
	{\begin{minipage}[t]{0.32\textwidth}
			\centering
			\includegraphics[width=0.95\textwidth]{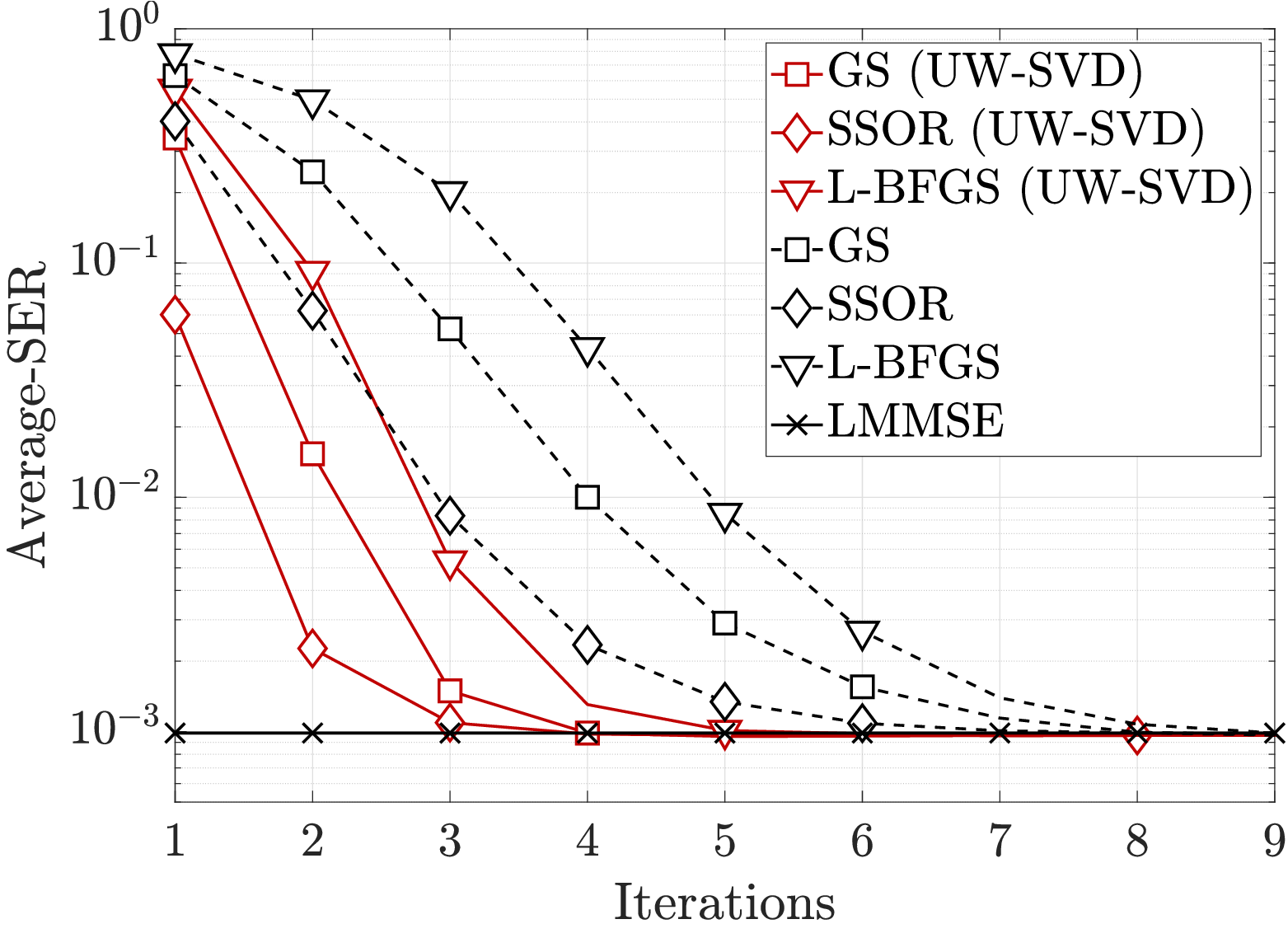}
			\label{fig04b}
	\end{minipage}}
	\subfigure[$\varrho = 0.8$; $\rho = 26.5$ $\mathrm{dB}$]{
		\begin{minipage}[t]{0.32\textwidth}
			\centering
			\includegraphics[width=0.95\textwidth]{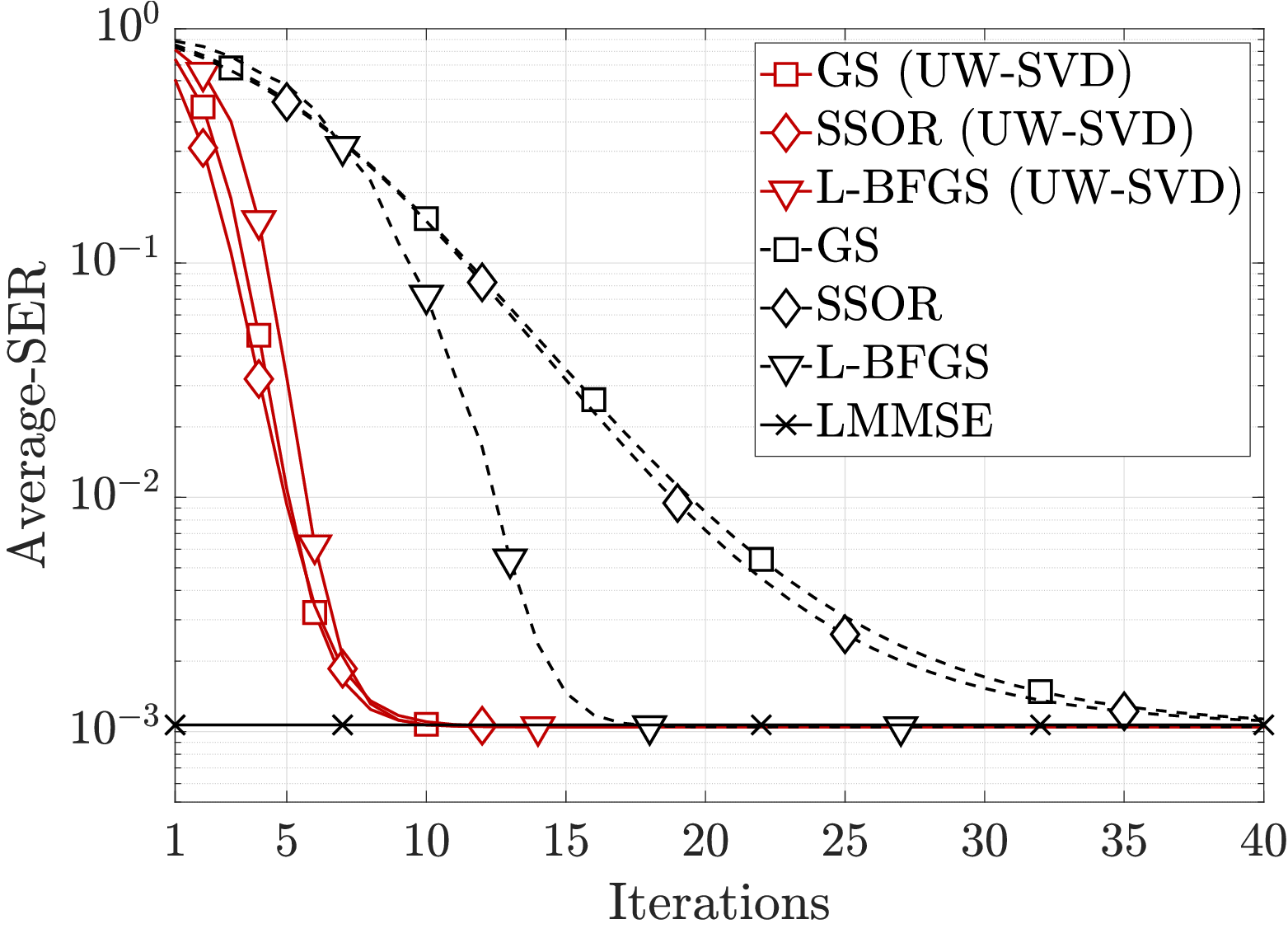}
			\label{fig04c}
	\end{minipage}}
	\caption{\label{fig04}Convergence comparison between different iterative algorithms in \textbf{Model 1}. $M = 256$; $K = 8$; $N_{\textsc{ue}} = 4$; $16$ QAM. UW-SVD-assisted algorithms can provide faster convergence compared to the corresponding existing algorithms, especially for correlated MIMO channels.}
	\vspace{-1.3em}
\end{figure*}

Fig. \ref{fig03} shows the results in the presence of LoS links, which can make the wireless channel more ill-conditioned.
In \textbf{Model 3} ($\varrho = 0$), it can be observed that $\mathrm{cond}(\mathbf{A}_{\textsc{zf}})$ is approximately $60$, meaning that $\mathbf{A}_{\textsc{zf}}$ is ill-conditioned.
Moreover, it will become even worse as the spatial correlation becomes higher, e.g., $\mathrm{cond}(\mathbf{A}_{\textsc{zf}}) \approx 600$ when $\varrho = 0.8$.
In addition, $\mathrm{cond}(\mathbf{A}_{\textsc{lmmse}})$ is smaller than $\mathrm{cond}(\mathbf{A}_{\textsc{zf}})$ due to the regularization term.
Similar observations can also be found in \textbf{Model 4}.
Moreover, $\mathrm{cond}(\mathbf{A})$ in \textbf{Model 4} can be well-conditioned with a probability of about $0.2$.
This is because this model allows the mixture of LoS/NLoS links, and the randomly generated channel matrix could be in a fully NLoS state with a certain probability.
This also leads to higher CDF fluctuations for $\mathrm{cond}(\mathbf{A})$ in \textbf{Model 4} than in \textbf{Model 3}.
In contrast, the fluctuations of $\mathrm{cond}(\mathbf{\Phi})$ are very small, and the value of $\mathrm{cond}(\mathbf{\Phi})$ is close to that of i.i.d. Rayleigh fading channels.
This implies that UW-SVD-assisted iterative methods can maintain consistently fast convergence even in the presence of increased intra-user interference.

\begin{figure*}[t]
	\hspace{1em}
	\subfigure[\textbf{Model 2}; $16$ QAM; $\varrho = 0.5$]{
		\begin{minipage}[t]{0.245\textwidth}
			\centering
			\includegraphics[height=5.3cm]{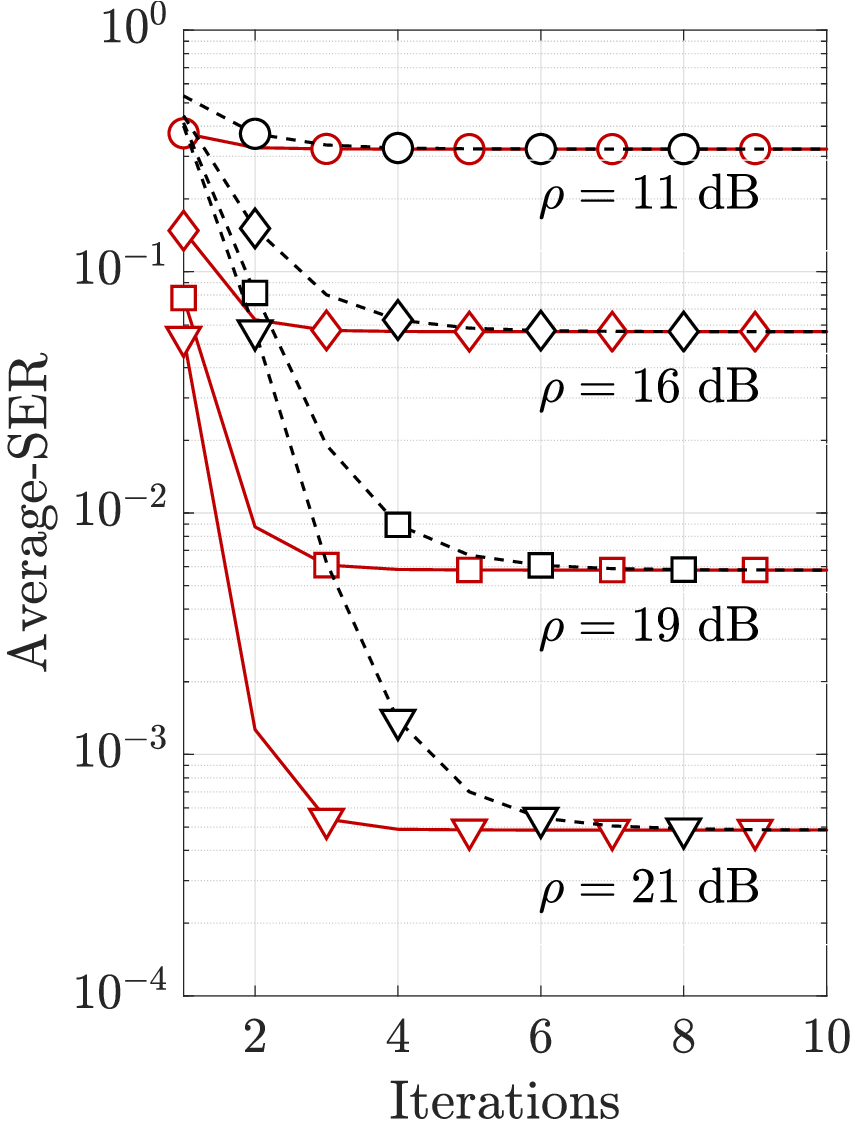}
			\label{fig05a}
	\end{minipage}}
	\subfigure[\textbf{Model 2}; $16$ QAM; $\varrho = 0.8$]{
		\begin{minipage}[t]{0.22\textwidth}
			\centering
			\includegraphics[height=5.3cm]{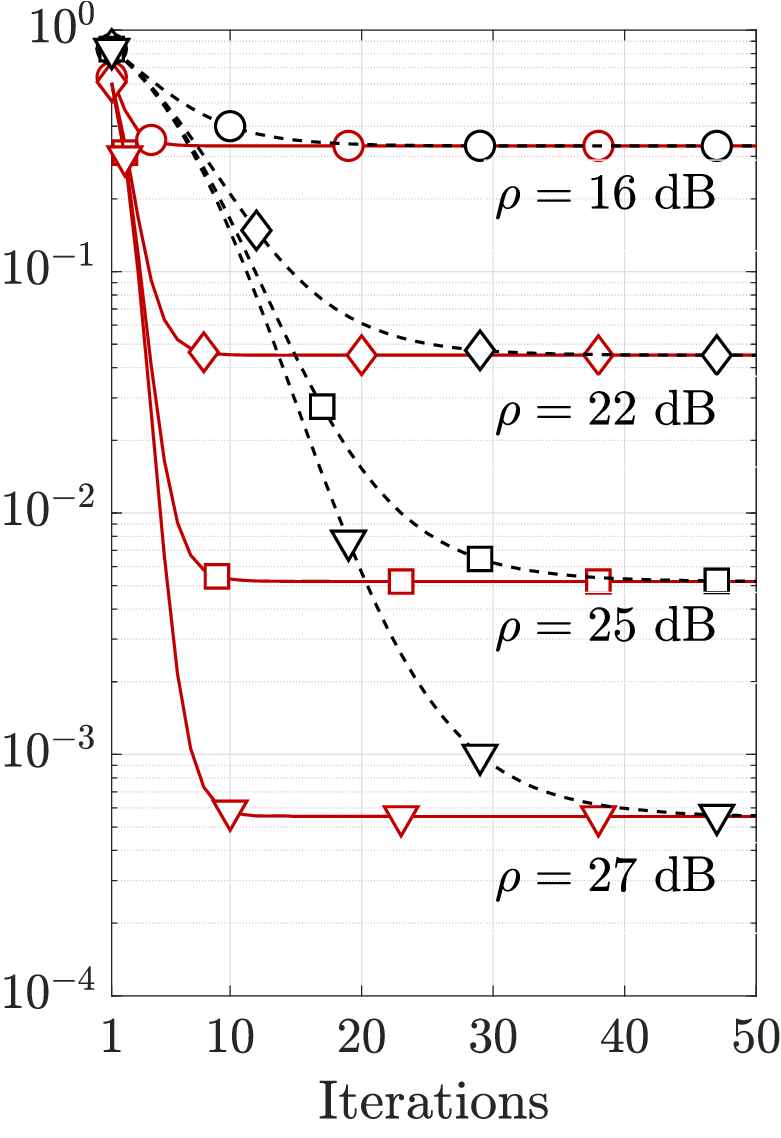}
			\label{fig05b}
	\end{minipage}}
	\subfigure[\textbf{Model 4}; $64$ QAM; $\varrho = 0.5$]{
		\begin{minipage}[t]{0.22\textwidth}
			\centering
			\includegraphics[height=5.3cm]{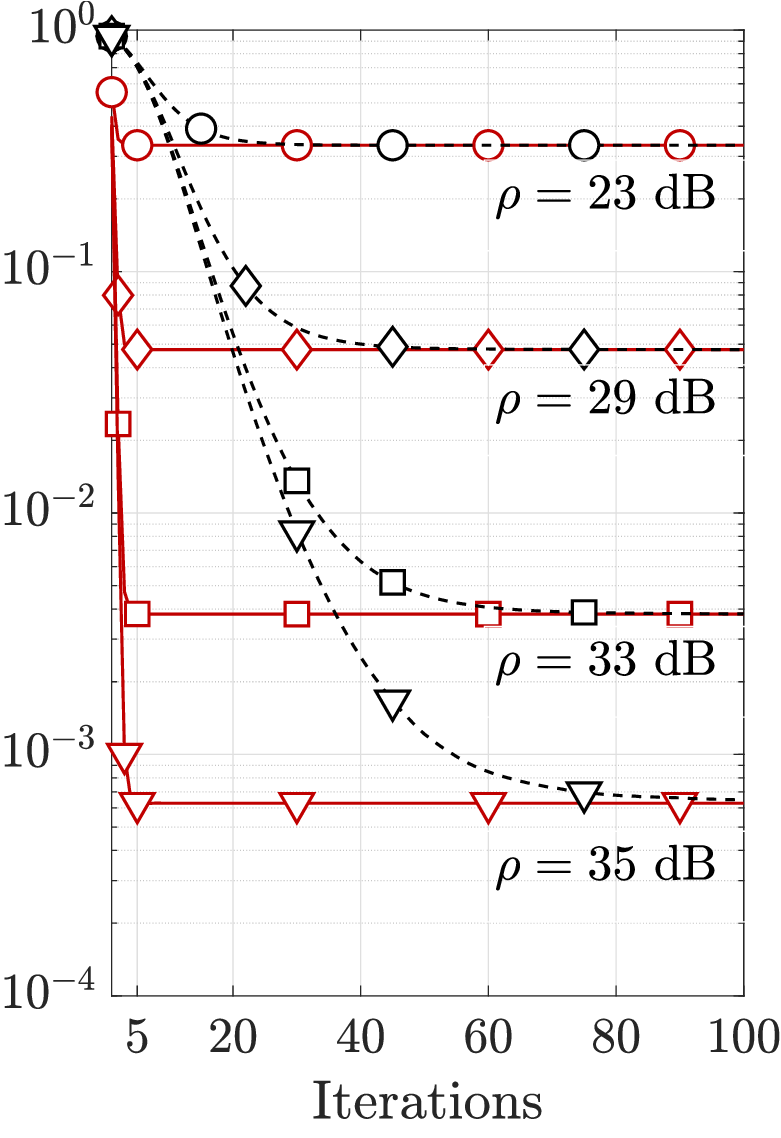}
			\label{fig05c}
	\end{minipage}}
	\subfigure[\textbf{Model 4}; $64$ QAM; $\varrho = 0.8$]{
		\begin{minipage}[t]{0.22\textwidth}
			\centering
			\includegraphics[height=5.3cm]{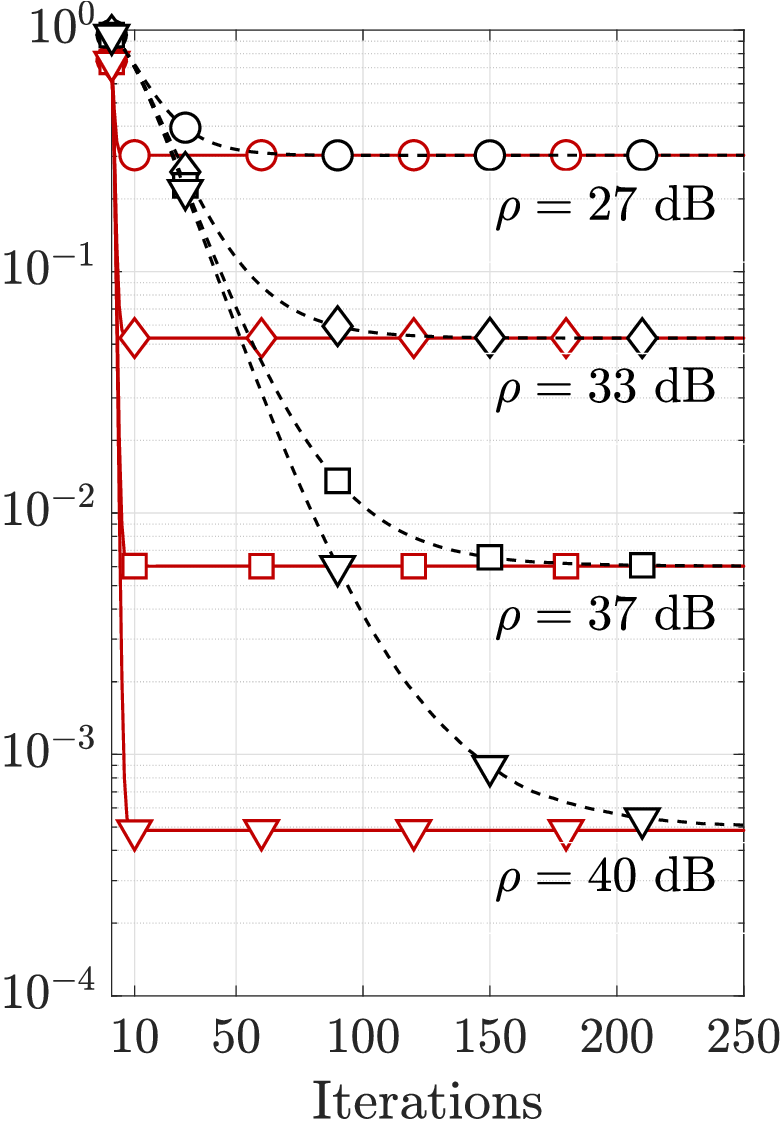}
			\label{fig05d}
	\end{minipage}}
	\caption{\label{fig05}Convergence comparison between UW-SVD-assisted SSOR (red lines) and SSOR (black lines) methods converging to LMMSE detection performance; $M = 256$; $K = 8$; $N_{\textsc{ue}} = 4$. It is shown that SSOR (UW-SVD) converges faster than SSOR at all different SNR levels.}
	\vspace{-1em}
\end{figure*}

\begin{figure*}[t]
	\centering
	\subfigure[$\varpi = 20$ $\mathrm{dB}$]{
	\begin{minipage}[t]{0.32\textwidth}
		\centering
		\includegraphics[width=0.95\textwidth]{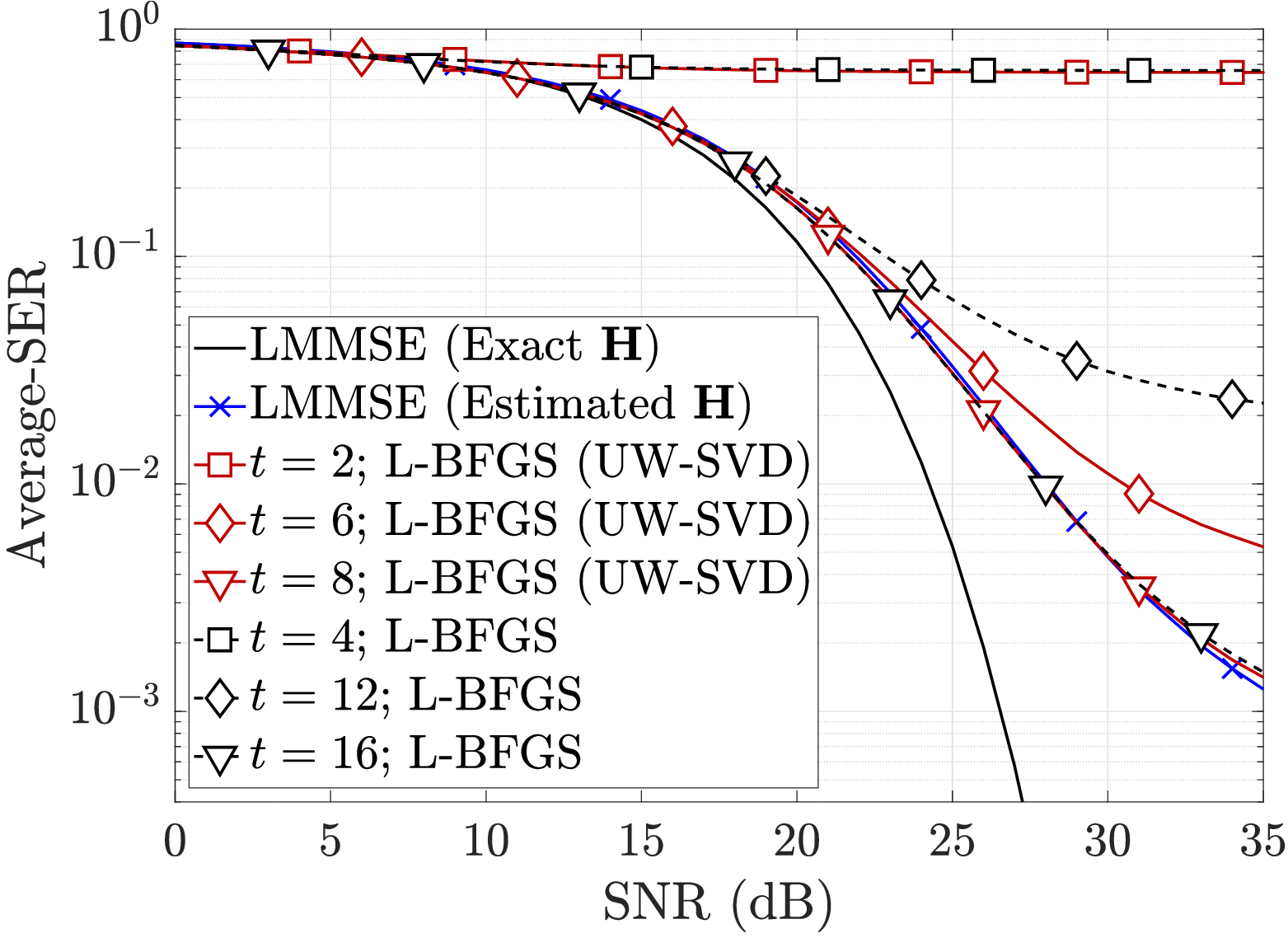}
	\end{minipage}}
	\subfigure[$\varpi = 15$ $\mathrm{dB}$]{
	\begin{minipage}[t]{0.32\textwidth}
		\centering
		\includegraphics[width=0.95\textwidth]{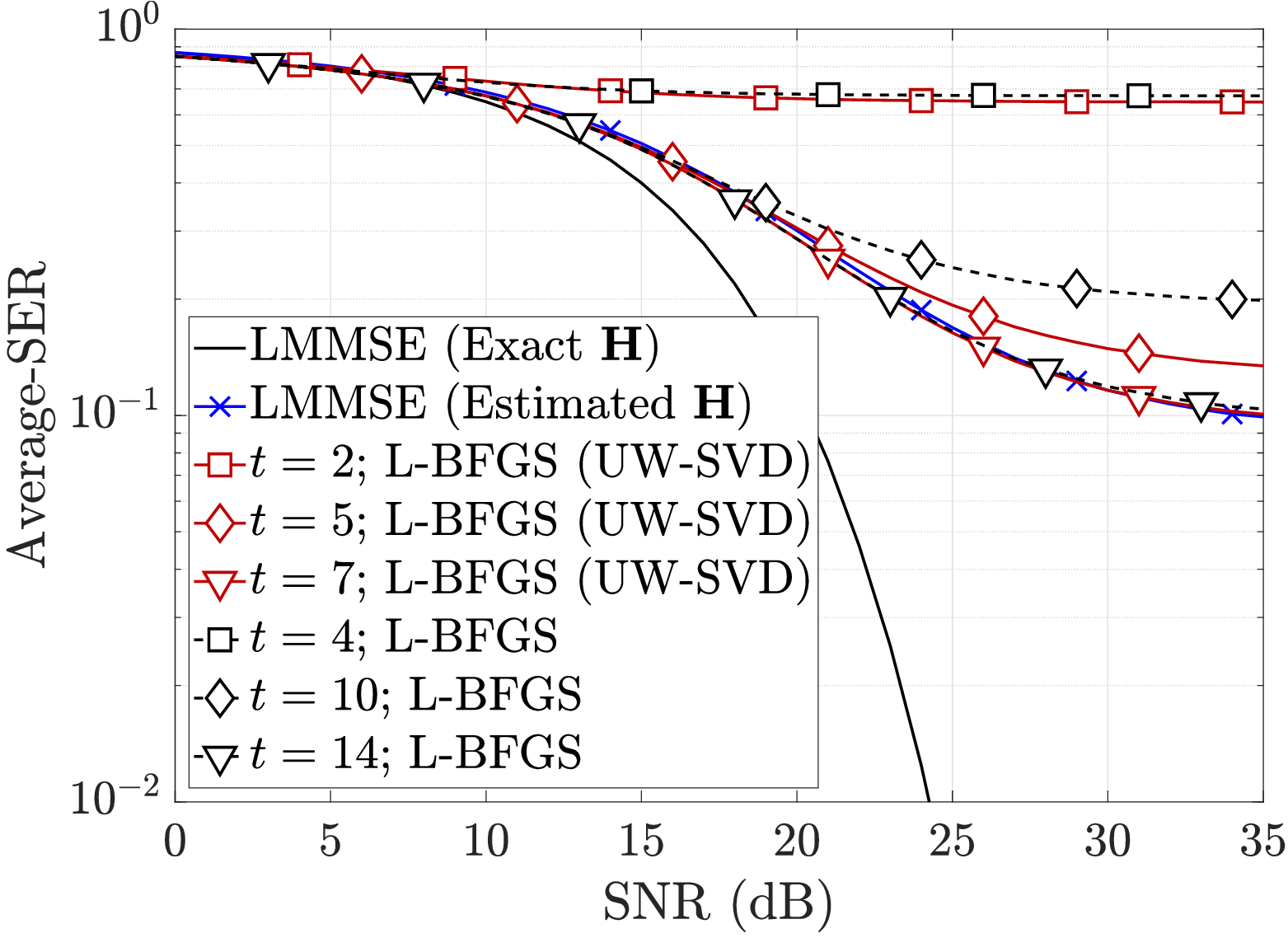}
	\end{minipage}}
	\subfigure[$\varpi = 10$ $\mathrm{dB}$]{
	\begin{minipage}[t]{0.32\textwidth}
		\centering
		\includegraphics[width=0.95\textwidth]{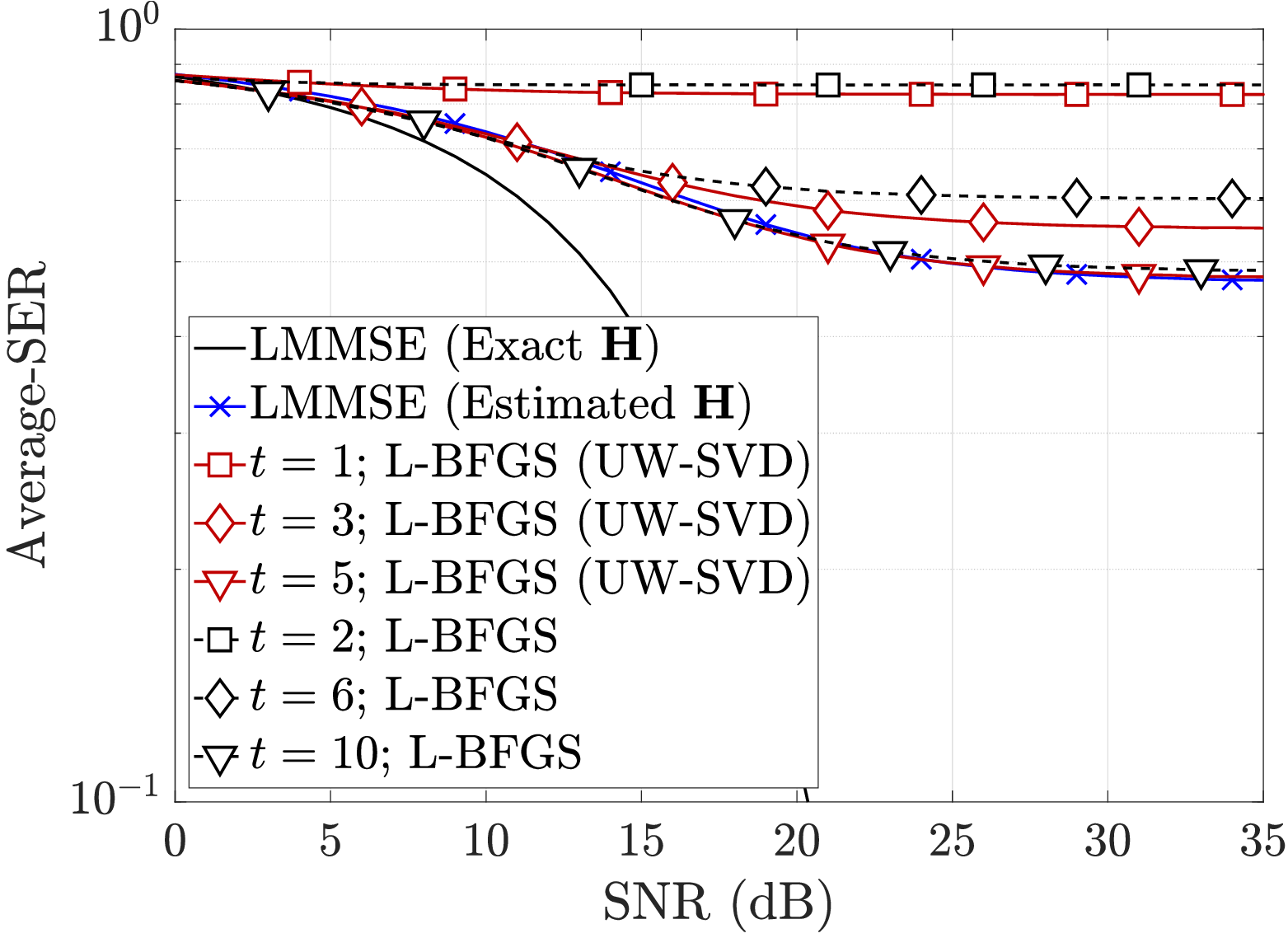}
	\end{minipage}}
	\caption{\label{fig06}Convergence comparison between L-BFGS (UW-SVD) and L-BFGS method with channel estimation error; \textbf{Model 2}; $M = 256$; $K = 8$; $N_{\textsc{ue}} = 4$. $16$ QAM; $\varrho = 0.2$. The proposed UW-SVD method can double the convergence speed of L-BFGS method for all three levels of channel estimation error.}
	\vspace{-1em}
\end{figure*}

\begin{figure}[t]
	\centering
	\subfigure{
		\begin{minipage}[t]{0.49\textwidth}
			\centering
			\includegraphics[width=0.76\textwidth]{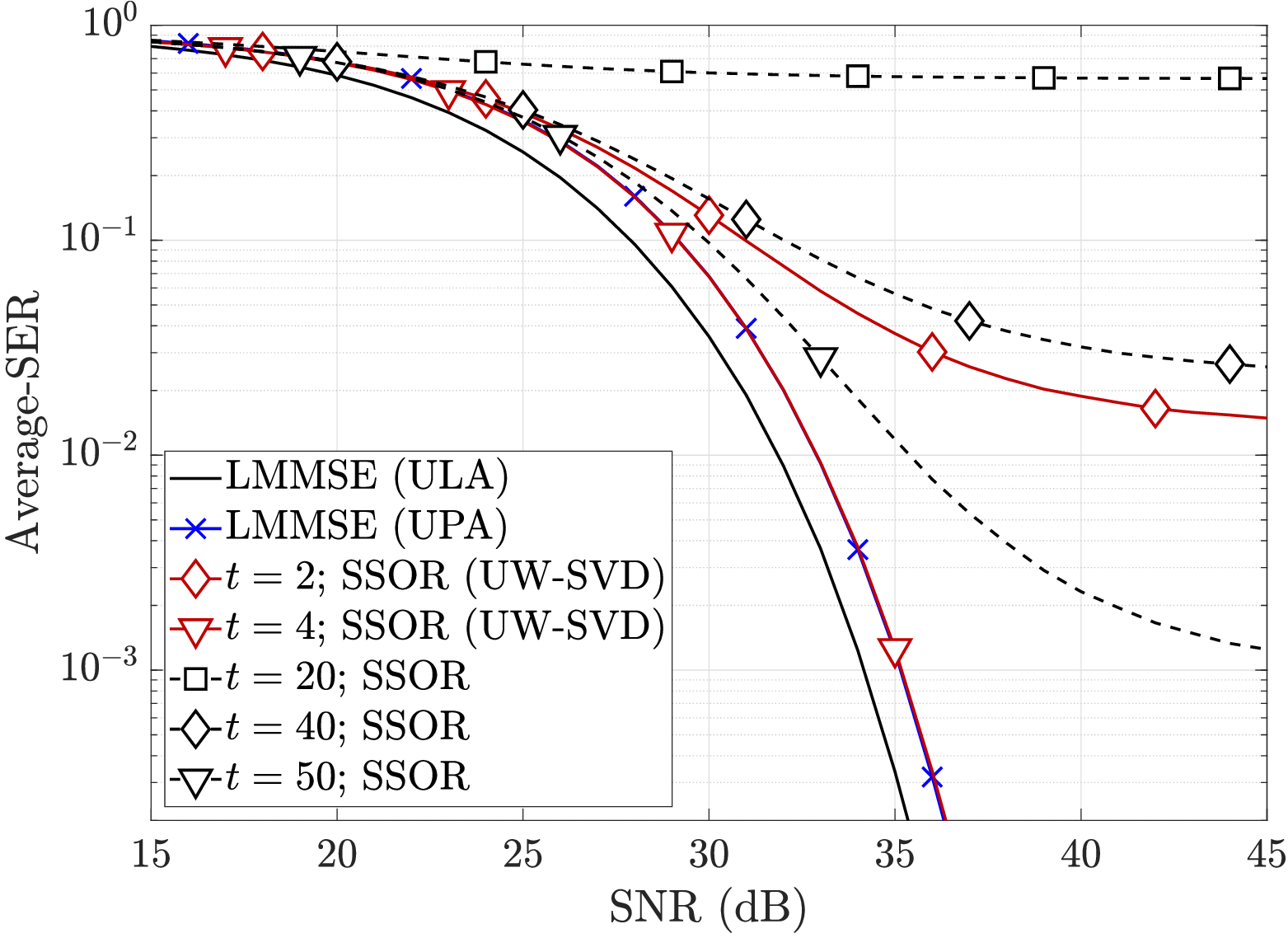}
	\end{minipage}}
	\caption{\label{fig07}Convergence comparison between SSOR (UW-SVD) and SSOR method for UPA configuration; \textbf{Model 3}; $64$ QAM; $\varrho = 0.5$.}
	\vspace{-1em}
\end{figure}

\textbf{Experiment 2: }
The objective is to demonstrate that the proposed UW-SVD-assisted iterative algorithms converge faster than corresponding existing algorithms in (ELAA-)MIMO systems.
In this experiment, four figures (i.e., Fig. \ref{fig04} - Fig. \ref{fig07}) are presented to highlight the advantages of the proposed UW-SVD method from different perspectives.
In Fig. \ref{fig04}, the convergence comparison between different iterative algorithms at high SNRs is shown.
It shows the average symbol error rate (SER) over the iterations in \textbf{Model 1}, considering three correlation levels.
For the case $\varrho = 0$, it can be seen that the proposed UW-SVD method can slightly accelerate the convergence of existing algorithms.
However, it is worth noting that the advantage of the proposed UW-SVD method becomes more apparent as the correlation becomes larger.
This is consistent with the numerical results in \textbf{Experiment 1}, and this figure indicates that the proposed UW-SVD method can accelerate the current iterative algorithm in conventional massive MIMO channels.

In Fig. \ref{fig05}, we aim to demonstrate the advantages of UW-SVD at different SNRs, by using SSOR that converges to LMMSE detection performance as an example.
Two ELAA channels (i.e., \textbf{Model 2} and \textbf{Model 4}) are considered in the figure, each with two correlation factors (i.e., $\varrho = 0.5$ and $\varrho = 0.8$).
As can be seen in each sub-figure, the advantage of UW-SVD diminishes with decreasing SNR.
This is consistent with our theoretical analysis in section \ref{sec04}.
However, it is worth noting that UW-SVD-assisted SSOR still converges significantly faster than the original SSOR method even at lower SNRs.
For instance, in Fig. \ref{fig05b}, when the SNR is $16$ $\mathrm{dB}$, the original SSOR method requires approximately $20$ iterations to converge in \textbf{Model 2} using $16$ QAM. 
However, with the assistance of UW-SVD, convergence is achieved in just $4$ iterations under the same system configuration.
As shown in Figs. \ref{fig05c} and \ref{fig05d}, the original SSOR method requires tens or even hundreds of iterations to achieve the LMMSE detection performance, even in low SNR scenarios.
In contrast, the UW-SVD-assisted SSOR method only requires fewer than $10$ iterations to converge.
This figure implies that the proposed UW-SVD method can accelerate the convergence of iterative algorithms at different SNR levels.

In Fig. \ref{fig06}, the objective is to demonstrate the robustness of the proposed UW-SVD method when channel estimation error is considered.
Let us consider the conventional LS channel estimation approach, and the estimated channel matrix is given by \cite{fowc}
\begin{equation}
	\widehat{\mathbf{H}} = \mathbf{H} + \mathbf{Z},
\end{equation}
where $\widehat{\mathbf{H}}$ denotes the estimated channel matrix, and $\mathbf{Z}$ is the AWGN matrix.
The ratio (denoted by $\varpi$) between the power of channel and noise elements is set to be $10$, $15$, and $20$ $\mathrm{dB}$ for the three sub-figures, respectively.
The MIMO channel is set to be \textbf{Model 2} with $\varrho = 0.2$ and the modulation scheme is set to be $16$ QAM.
It can be observed that the LMMSE detector with channel estimation error can only provide sub-optimal detection performance.
Therefore, all the iterative algorithms will only converge to this sub-optimal detection performance.
The proposed UW-SVD method consistently accelerate the convergence of the L-BFGS method by a factor of two, irrespective of the level of channel estimation error.
More specifically, the UW-SVD-assisted L-BFGS method converges within $10$, $7$, and $5$ iterations for $\varpi = 20$ $\mathrm{dB}$, $15$ $\mathrm{dB}$, and $10$ $\mathrm{dB}$, respectively.
In contrast, the original L-BFGS method requires $20$, $14$, and $10$ iterations to converge for the same respective levels of $\varpi$.
These results show that UW-SVD can improve the convergence speed of L-BFGS by approximately two times for different levels of channel estimation error.

In Fig. \ref{fig07}, the objective is to show that UW-SVD can accelerate current iterative algorithms in another type of ELAA, i.e., UPA.
The UPA is configured with $M = 16 \times 16$ antennas.
The simulation results depicted in Fig. \ref{fig07} suggest a performance degradation for the LMMSE detector in UPA compared to its performance in ULA.
The reason for this is that UPA antennas are more tightly distributed, resulting in higher spatial correlations.
In this figure, we utilize the SSOR method that converges to the LMMSE detection performance to demonstrate the advantages of the proposed UW-SVD method.
It is noteworthy that the UW-SVD-assisted SSOR method converges faster than the original SSOR method across different SNR levels. 
For instance, the original SSOR method necessitates over $50$ iterations to converge, and it requires more than $20$ iterations even at relatively low SNR. 
Conversely, the SSOR method assisted by UW-SVD achieves convergence in only $4$ iterations at all SNR levels.

{\bf Experiment 3:}
\begin{figure}[t]
	\subfigure[\textbf{Model 3}; Convolutional code; $16$ QAM; $K = 8$]{
	\begin{minipage}[t]{0.49\textwidth}
		\label{fig21312211a}
		\centering
		\includegraphics[width=0.76\textwidth]{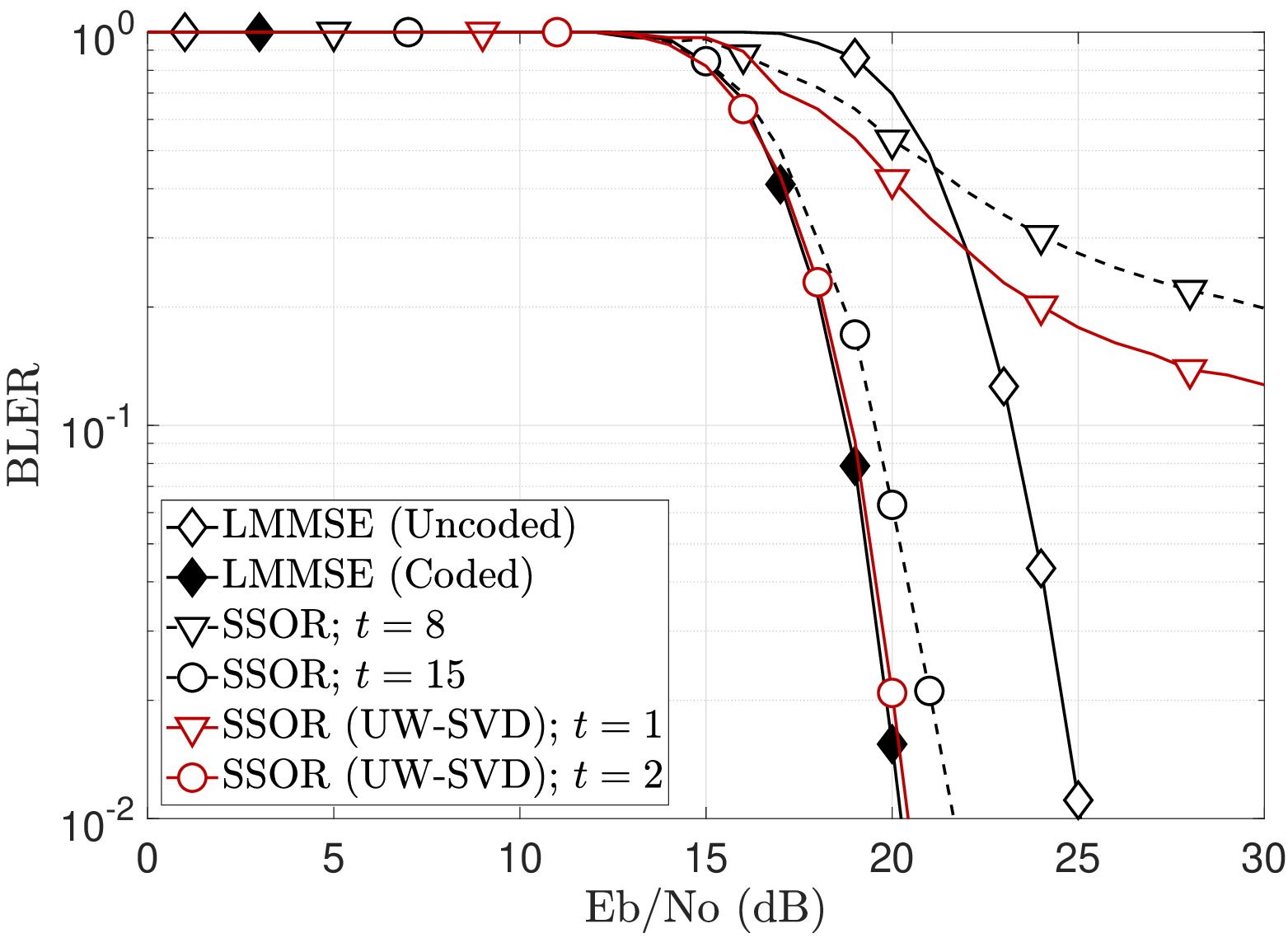}
		\vspace{0.5em}		
	\end{minipage}}
	\subfigure[\textbf{Model 4}; Polar code; $64$ QAM; $K = 16$]{
	\begin{minipage}[t]{0.49\textwidth}
		\label{fig21312211b}
		\centering
		\includegraphics[width=0.76\textwidth]{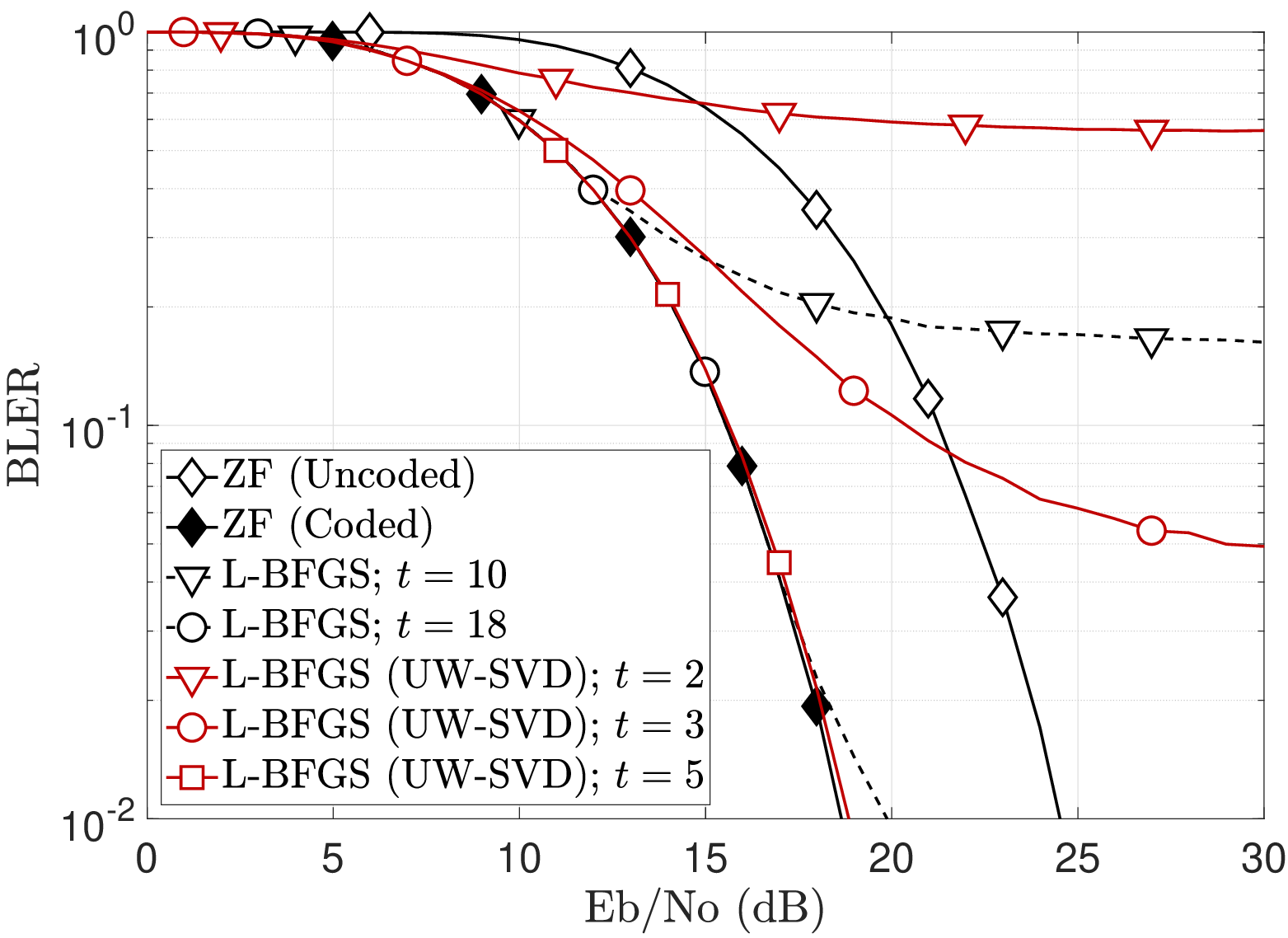}
		\vspace{0.5em}		
	\end{minipage}}		
	\caption{\label{fig21312211}Convergence comparison of iterative algorithms in ELAA-MIMO systems considering channel coding; $M = 256$; $N_{\textsc{ue}} = 4$. The UW-SVD-assisted SSOR and L-BFGS methods achieve convergence ten and five times faster than the original SSOR and L-BFGS methods, respectively.}
	\vspace{-1em}
\end{figure}
The objective of this experiment is to demonstrate that, with channel coding, the UW-SVD can still significantly accelerate the convergence of current algorithms.
Two coding schemes are considered: $1/2$ convolutional code with a codeword length of $200$ bits, and $1/4$ polar code with a codeword length of $1,024$ bits.
The decoding schemes are Viterbi decoder and successive cancellation list for convolutional code and polar code, respectively.
The modulation schemes are $16$ QAM and $64$ QAM for convolutional and polar codes, respectively.
In addition, the performance metric is set to block error rate (BLER) versus Eb/No.
As shown in Fig. \ref{fig21312211}, the performance gap between uncoded and coded systems is approximately $6$ $\mathrm{dB}$ for both channel models.
In Fig. \ref{fig21312211a}, the UW-SVD-assisted SSOR method converges to the LMMSE detection performance in only $2$ iterations, while the SSOR method requires over $15$ iterations to achieve the same level of convergence.
In coded MIMO systems, the improvements achieved by UW-SVD for SSOR remain comparable to those observed in uncoded MIMO systems.
Moreover, as shown in Fig. \ref{fig21312211b}, UW-SVD-assisted L-BFGS methods can achieve ZF detection performance within three iterations, while the standard L-BFGS algorithm requires over $15$ iterations to achieve the same level of performance.
Together with the results in \textbf{Experiment 2}, it can be claimed that UW-SVD can significantly accelerate the convergence of current algorithms by up to ten times, in both uncoded and coded MIMO systems.

\section{Conclusion} \label{sec06}
In this paper, we propose the UW-SVD method to accelerate the convergence of current iterative algorithms for spatially correlated (ELAA-)MIMO channels.
The results demonstrate that the UW-SVD-assisted algorithms achieve convergence up to more than ten times faster compared to the corresponding current algorithms in both coded and uncoded systems.
The core principle is to perform SVD on each user's sub-channel matrix, transforming the original MIMO signal model into an e-signal model. 
For this e-signal model, we develop e-ZF and e-LMMSE detectors with detection performance proven to be equivalent to ZF and LMMSE detectors for the original model.
Crucially, it is shown that the e-channel matrix exhibits a significantly better condition number than the original MIMO channel matrix, when considering the channel spatial correlation or non-stationarity or both.
By applying current iterative algorithms to iteratively invert the better-conditioned e-channel matrix, followed by a post-processing step to recover the transmitted signals, remarkable convergence acceleration is achieved.

\appendices
\section{Proof of \textit{Theorem \ref{thm03}}} \label{appThm03}
According to \textit{Property} \ref{lem01}, it is straightforward that
\begin{equation}
\lim\limits_{M \rightarrow \infty} \mathbf{\Sigma}_{k} = \mathbf{I}, \quad \forall k.
\end{equation}
Hence, we have
\begin{equation} \label{eqn16322508}
\lim\limits_{M \rightarrow \infty} \mathbf{\Sigma} = \mathrm{diag}
(\mathbf{\Sigma}_{1}, \ldots, \mathbf{\Sigma}_{K})  = \mathbf{I}.
\end{equation}
Plugging \eqref{eqn09470308} into $\mathbf{A}_{\textsc{zf}}$ yields
\begin{equation} \label{eqn22092711}
	\mathbf{A}_{\textsc{zf}}  = \mathbf{V} \mathbf{\Sigma} \mathbf{\Psi}^{H} \mathbf{\Psi} \mathbf{\Sigma} \mathbf{V}^{H}.
\end{equation}
Plugging \eqref{eqn16322508} into \eqref{eqn22092711} yields
\begin{equation} \label{eqn22122711}
	\lim\limits_{M \rightarrow \infty} \mathbf{A}_{\textsc{zf}}
	= \mathbf{V} \mathbf{\Psi}^{H} \mathbf{\Psi} \mathbf{V}^{H}.
\end{equation}
Plugging $\mathbf{\Phi_{\textsc{zf}}} = \mathbf{\Psi}^{H}\mathbf{\Psi}$ into \eqref{eqn22122711} yields
\begin{equation} \label{eqn22132711}
	\lim\limits_{M \rightarrow \infty} \mathbf{A}_{\textsc{zf}}
	= \mathbf{V} \mathbf{\Phi}_{\textsc{zf}} \mathbf{V}^{H}.
\end{equation}
Given that $\mathbf{V}$ is a unitary matrix, it does not change the condition number of the matrix being multiplied.
Hence, \eqref{eqn22142711a} in \textit{Theorem} \ref{thm03} is proved.
Similarly, plugging \eqref{eqn09470308} and \eqref{eqn16322508} into $\mathbf{A}_{\textsc{lmmse}}$ yields
\begin{IEEEeqnarray}{ll}
	\lim\limits_{M \rightarrow \infty} \mathbf{A}_{\textsc{lmmse}}
	&= \mathbf{V} \mathbf{\Psi}^{H} \mathbf{\Psi} \mathbf{V}^{H} + \rho^{-1} \mathbf{I}, \nonumber \\
	&= \mathbf{V}  (\mathbf{\Psi}^{H} \mathbf{\Psi} + \rho^{-1} \mathbf{I}) \mathbf{V}^{H}. \label{eqn10121311}
\end{IEEEeqnarray}
According to \eqref{eqn16322508}, $\mathbf{\Phi}_{\textsc{lmmse}}$ in \eqref{eqn17011811} can be expressed as follows
\begin{equation} \label{eqn09571311}
	\lim\limits_{M \rightarrow \infty} \mathbf{\Phi}_{\textsc{lmmse}} = \mathbf{\Psi}^{H}\mathbf{\Psi} + \rho^{-1} \mathbf{I}.
\end{equation}
Plugging \eqref{eqn09571311} into \eqref{eqn10121311} yields
\begin{equation}
	\lim\limits_{M \rightarrow \infty} \mathbf{A}_{\textsc{lmmse}} = \mathbf{V} \mathbf{\Phi}_{\textsc{lmmse}} \mathbf{V}^{H}.
\end{equation}
Together with \eqref{eqn22132711}, \textit{Theorem} \ref{thm03} is proved.

\section{Proof of \textit{Lemma \ref{lem02}}} \label{appLem02}
According to the assumption \textit{A1}, the correlation matrix $\mathbf{R}_{\textsc{ue}}$ is a block diagonal matrix.
This indicates that $\sqrt{\mathbf{R}_{\textsc{ue}}}$ is also a block diagonal matrix, and it can be expressed as follows
\begin{equation}
	\sqrt{\mathbf{R}_{\textsc{ue}}} = \mathrm{diag} \bigg(\sqrt{\mathbf{R}_{\textsc{ue}}^{1, 1}}, \dots, \sqrt{\mathbf{R}_{\textsc{ue}}^{K, K}}\bigg).
\end{equation}
Hence, the sub-channel matrix of the $k^{th}$ user can be expressed by $\mathbf{H}_{k} = \mathbf{\Omega}_{k} \sqrt{\mathbf{R}_{\textsc{ue}}^{k, k}}$, resulting in
\begin{equation} \label{eqn18502502}
	\mathbf{H}_{k}^{H} \mathbf{H}_{j} = \sqrt{\mathbf{R}_{\textsc{ue}}^{k, k}} \mathbf{\Omega}_{k}^{H}  \mathbf{\Omega}_{j} \sqrt{\mathbf{R}_{\textsc{ue}}^{j, j}},
\end{equation}
where $\mathbf{\Omega}_{k} \in \mathbb{C}^{M \times N_{k}}$ represents the i.i.d. Rayleigh distributed matrix.
According to \textit{Property} \ref{lem01}, we have the following
\begin{equation} \label{eqn13120203}
	\lim\limits_{M \rightarrow \infty} \mathbf{\Omega}_{k}^{H}  \mathbf{\Omega}_{j}  = \mathbf{0}, \quad \forall k \neq j.
\end{equation}
Applying (\ref{eqn13120203}) into (\ref{eqn18502502}), \textit{Lemma} \ref{lem02} are therefore obtained.

\section{Proof of \textit{Theorem \ref{thm04}}} \label{appThm04}
Plugging \eqref{eqn21381310} into $\mathbf{A}_{\textsc{zf}}$ yields
\begin{equation}
\mathbf{A}_{\textsc{zf}} = [\mathbf{H}_{1}^{H}, \dots , \mathbf{H}_{K}^{H}]^{T}[\mathbf{H}_{1}, ... ,\mathbf{H}_{K}].
\end{equation}
According to \textit{A2}, it can be found that all the non-diagonal parts of $\mathbf{A}_{\textsc{zf}}$ are $\mathbf{0}$.
Hence, we have the following
\begin{equation}\label{eqn21012011}
	\mathbf{A}_{\textsc{zf}} = \mathrm{diag}(\mathbf{H}_{1}^{H}\mathbf{H}_{1}, \dots, \mathbf{H}_{K}^{H}\mathbf{H}_{K}), 
\end{equation}
which indicate that $\mathbf{A}_{\textsc{zf}}$ is a block diagonal matrix.
Therefore, $\mathrm{cond}(\mathbf{A}_{\textsc{zf}})$ should not be smaller than the condition number of any of its blocks, i.e.,
\begin{equation} \label{eqn11151611}
	\mathrm{cond}(\mathbf{A}_{\textsc{zf}}) \geq \max\{\mathrm{cond}(\mathbf{H}_{k}^{H} \mathbf{H}_{k})\}.
\end{equation}
Since the intra-user channel columns are correlated, it is clear that $\mathrm{cond}(\mathbf{H}_{k}^{H}\mathbf{H}_{k}) > 1$, and we have the following
\begin{equation} \label{eqn21232011}
\mathrm{cond}(\mathbf{A}_{\textsc{zf}}) > 1.
\end{equation}
Also, given \textit{A1}, performing SVD on $\mathbf{H}_{k}$ and $\mathbf{H}_{j}$ yields
\begin{equation} \label{eqn11061611}
	\mathbf{H}_{k}^{H}\mathbf{H}_{j} = \mathbf{V}_{k}\mathbf{\Sigma}_{k} \mathbf{U}_{k}^{H}\mathbf{\mathbf{U}}_{j}\mathbf{\Sigma}_{j}\mathbf{V}_{j}^{H} = \mathbf{0}, \quad \forall k \neq j.
\end{equation}
Right multiplying $\mathbf{\Sigma}_{k}^{-1}\mathbf{V}_{k}^{H}$ and left multiplying $\mathbf{V}_{j}\mathbf{\Sigma}_{j}^{-1}$ on \eqref{eqn11061611} yields
\begin{equation}
	\mathbf{U}_{k}^{H}\mathbf{U}_{j} = \mathbf{0}, \quad \forall k \neq j.
\end{equation}
Similar to that of $\mathbf{A}_{\textsc{zf}}$, i.e., \eqref{eqn21012011}, $\mathbf{\Phi}_{\textsc{zf}}$ can also be expressed as follows
\begin{equation}  \label{eqn11131611}
	\mathbf{\Phi}_{\textsc{zf}} = \mathrm{diag}(\mathbf{U}_{1}^{H}\mathbf{U}_{1}, \dots, \mathbf{U}_{K}^{H}\mathbf{U}_{K}),
\end{equation}
which indicates that $\mathbf{\Psi}_\textsc{zf} = \mathbf{I}$ with condition number $1$, since $\mathbf{U}_{k}, \forall k$ is a unitary matrix.
Together with \eqref{eqn21232011}, \eqref{eqn10352811} in \textit{Theorem} \ref{thm04} is therefore obtained.
	
\section{Proof of \textit{Theorem \ref{thm05}}} \label{appThm05}
According to \eqref{eqn10531708}, $\mathbf{A}_{\textsc{lmmse}}$ can be expressed as follows
\begin{equation}
	\mathbf{A}_{\textsc{lmmse}} = \mathbf{A}_{\textsc{zf}} + \rho^{-1} \mathbf{I}.
\end{equation}
Therefore, $\mathrm{cond}(\mathbf{A}_{\textsc{lmmse}})$ can be expressed as follows
\begin{equation} \label{eqn11431611}
	\mathrm{cond}(\mathbf{A}_{\textsc{lmmse}}) = \Bigg(\dfrac{\lambda_{\text{max}}(\mathbf{A}_{\textsc{zf}}) + \rho^{-1}}{\lambda_{\text{min}}(\mathbf{A}_{\textsc{zf}}) + \rho^{-1}}\Bigg).
\end{equation}
According to \eqref{eqn17011811}, $\mathbf{\Phi}_{\textsc{lmmse}}$ can be expressed as follows
\begin{equation}
	\mathbf{\Phi}_{\textsc{lmmse}} = \mathbf{\Phi}_{\textsc{zf}} + \rho^{-1} \mathbf{\Sigma}^{-2}.
\end{equation}
According to \eqref{eqn11131611} in \textit{Theorem} \ref{thm04}, we have the following
\begin{equation} \label{eqn11371611}
	\mathrm{cond}(\mathbf{\Phi}_{\textsc{lmmse}}) = \Bigg(\dfrac{1 + \rho^{-1}\lambda_{\text{max}}(\mathbf{\Sigma}^{-2})}{1 + \rho^{-1}\lambda_{\text{min}}(\mathbf{\Sigma}^{-2})}\Bigg).
\end{equation}
According to \eqref{eqn21012011}, $\mathbf{A}_{\textsc{zf}}$ is a block diagonal matrix.
Moreover, $\mathbf{\Sigma}$ contains the singular values of $\mathbf{H}_{k}, \forall k$, so that $\mathbf{\Sigma}^{2}$ contains the eigenvalues of every block in $\mathbf{A}_{\textsc{zf}}$.
Hence,  we have the following
\begin{equation} \label{eqn11351611}
	\lambda_{\text{max}}(\mathbf{\Sigma}^{-2}) = \lambda_{\text{min}}(\mathbf{A}_{\textsc{zf}})^{-1};
\end{equation}
\begin{equation} \label{eqn11361611}
	\lambda_{\text{min}}(\mathbf{\Sigma}^{-2}) = \lambda_{\text{max}}(\mathbf{A}_{\textsc{zf}})^{-1}.
\end{equation}
Plugging \eqref{eqn11351611} and \eqref{eqn11361611} into \eqref{eqn11371611} with some tidy up works yields
\begin{equation} \label{eqn11441611}
	\mathrm{cond}(\mathbf{\Phi}_{\textsc{lmmse}}) = \Bigg(\dfrac{\lambda_{\text{max}}(\mathbf{A}_{\textsc{zf}})}{\lambda_{\text{min}}(\mathbf{A}_{\textsc{zf}})}\Bigg) \Bigg(\dfrac{ \lambda_{\text{min}}(\mathbf{A}_{\textsc{zf}}) + \rho^{-1}}{\lambda_{\text{max}}(\mathbf{A}_{\textsc{zf}}) + \rho^{-1}}\Bigg).
\end{equation}
It is obvious that the left term in \eqref{eqn11441611} is $\mathrm{cond}(\mathbf{A}_{\textsc{zf}})$. 
Moreover, according to \eqref{eqn11431611}, $\mathrm{cond}(\mathbf{\Phi}_{\textsc{lmmse}})$ in \eqref{eqn11441611} can be expressed as follows
\begin{equation} \label{eqn11561611}
	\mathrm{cond}(\mathbf{\Phi}_{\textsc{lmmse}}) = \dfrac{\mathrm{cond}(\mathbf{A}_{\textsc{zf}}) }{\mathrm{cond}(\mathbf{A}_{\textsc{lmmse}})}.
\end{equation}
To obtain the condition under which \eqref{eqn11042811} in \textit{Theorem} \ref{thm05} holds, plugging \eqref{eqn11561611} into \eqref{eqn11042811} yields
\begin{equation} \label{eqn12001611}
	\mathrm{cond}(\mathbf{A}_{\textsc{zf}}) < \mathrm{cond}^{2}(\mathbf{A}_{\textsc{lmmse}}).
\end{equation}
Plugging \eqref{eqn11431611} into \eqref{eqn12001611}
\begin{equation}
	\Bigg(\dfrac{\lambda_{\text{max}}(\mathbf{A}_{\textsc{zf}})}{\lambda_{\text{min}}(\mathbf{A}_{\textsc{zf}})}\Bigg)  <  \Bigg(\dfrac{\lambda_{\text{max}}(\mathbf{A}_{\textsc{zf}}) + \rho^{-1}}{\lambda_{\text{min}}(\mathbf{A}_{\textsc{zf}}) + \rho^{-1}}\Bigg)^{2}.
\end{equation}
With some tidy-up works, this inequality holds if
\begin{equation} \label{eqn21302811}
	\rho > \dfrac{1}{\sqrt{\lambda_{\text{max}}(\mathbf{A}_{\textsc{zf}})\lambda_{\text{min}}(\mathbf{A}_{\textsc{zf}})}}.
\end{equation}
Given $\sigma_{x}^{2} = 1$, we have $\rho = 1 / \sigma_{z}^{2}$, and plugging it into \eqref{eqn21302811}, \eqref{eqn20201911} in \textit{Theorem} \ref{thm05} is therefore obtained.

\ifCLASSOPTIONcaptionsoff
\newpage
\fi

\bibliographystyle{myIEEEtran}
\bibliography{IEEEabrv,mMIMO}

\begin{IEEEbiography}[{\includegraphics[width=1in,height=1.25in,clip]{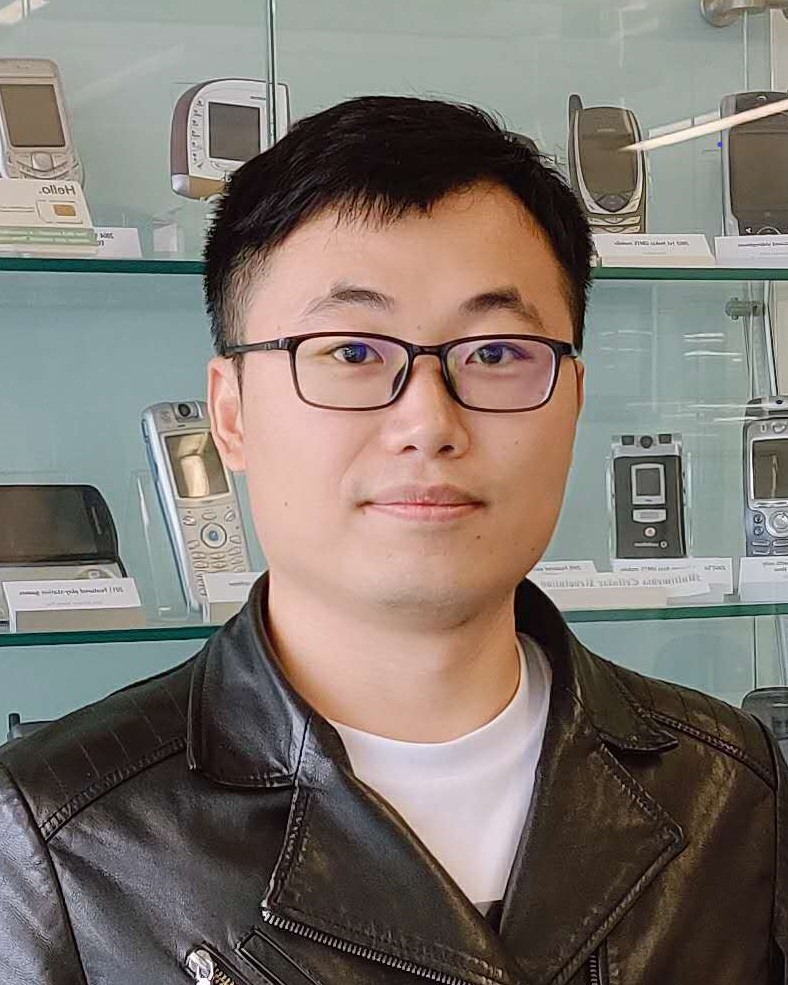}}]{Jiuyu Liu}
	(Graduate Student Member, IEEE) received the B.Eng. degrees in computer science and technology from the Xi'an JiaoTong-Liverpool University, Suzhou, China, as well as in computer science and electronic engineering from the University of Liverpool, U.K., in 2019. He received the M.Sc. degree in electronic engineering from the University of Surrey, U.K., in 2020. He is currently pursuing the Ph.D. degree in electronic engineering with the 5GIC \& 6GIC, Institute for Communication Systems, University of Surrey, U.K. His main research interests include multiple-input multiple-output, extremely large aperture array, spatially non-stationary channel modeling, stochastic process, digital signal processing.
\end{IEEEbiography}

\begin{IEEEbiography}[{\includegraphics[width=1in,height=1.25in,clip]{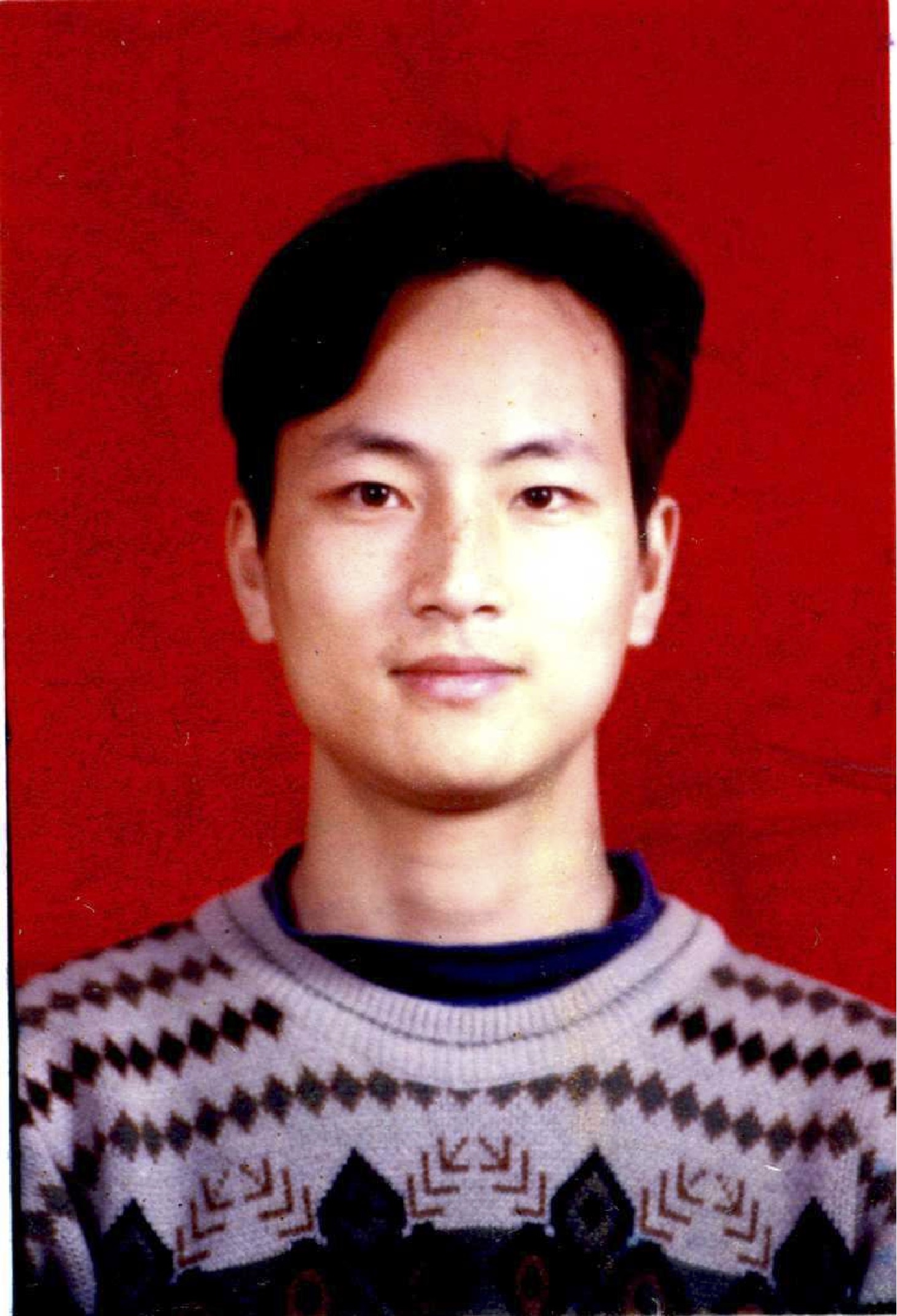}}]{Yi Ma}
	(Senior Member, IEEE) is currently a Chair Professor with the Institute for Communication Systems (ICS), University of Surrey, Guildford, U.K. He has authored or coauthored more than 200 peer reviewed IEEE journal articles and conference papers in the areas of deep learning, cooperative communications, cognitive radios, interference utilization, cooperative localization, radio resource allocation, multiple-input multiple-output, estimation, synchronization, and modulation and detection techniques. He holds ten international patents in the areas of spectrum sensing and signal modulation and detection. He has served as the Tutorial Chair for EuroWireless 2013, PIMRC 2014, and CAMAD 2015. He is the Co-Chair of the Signal Processing for Communications Symposium in ICC 2019. He was the Founder of the Crowd-Net Workshop in conjunction with ICC 2015, ICC 2016, and ICC 2017.
\end{IEEEbiography}

\begin{IEEEbiography}[{\includegraphics[width=1in,height=1.25in,clip]{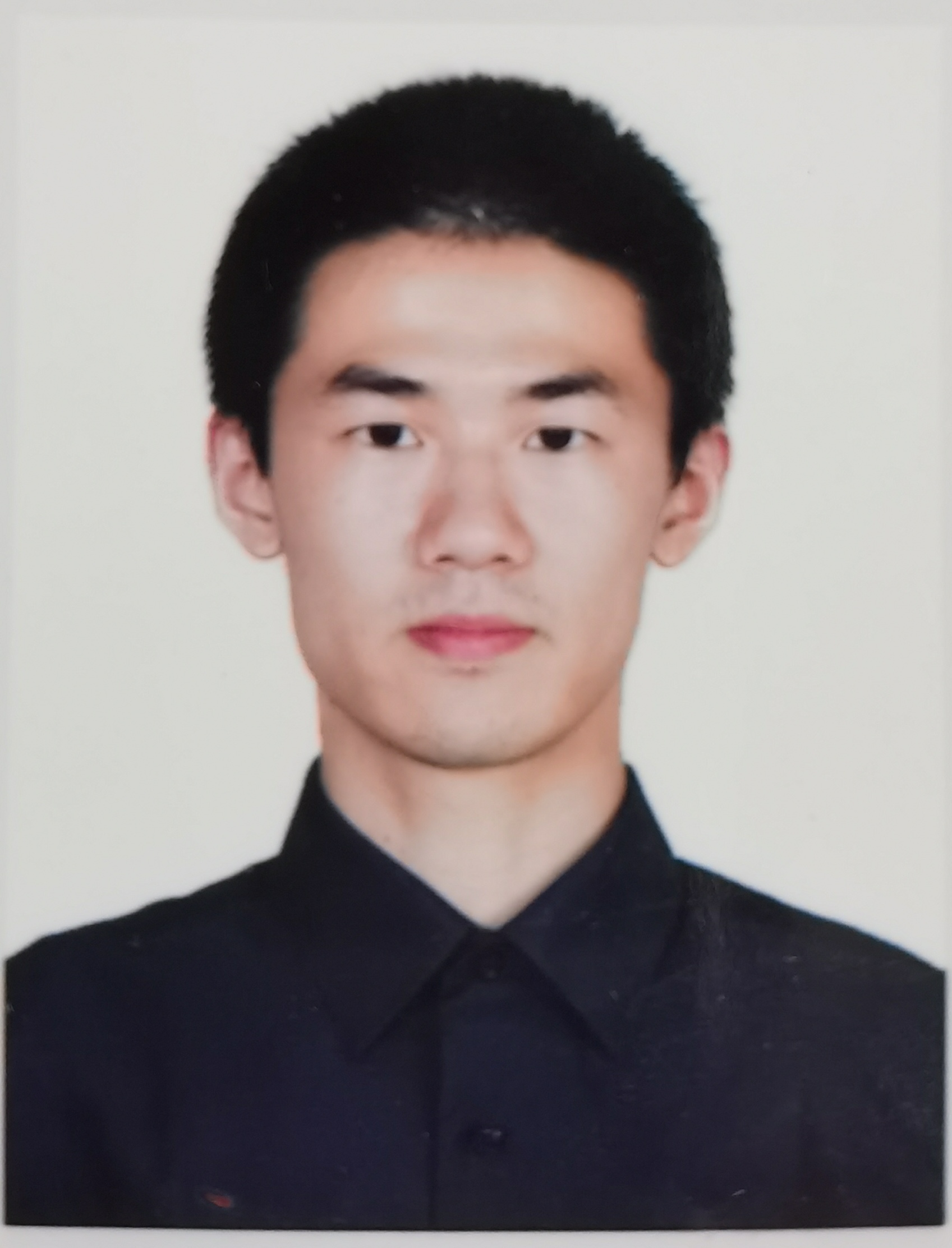}}]{Jinfei Wang}
	received his Ph.D. degree from the University of Surrey, U.K., in 2023. He is currently a Research Fellow at the 5GIC \& 6GIC, Institute for Communication Systems (ICS), University of Surrey, Guildford, U.K. His main research interests include: physical layer design of massive multiple-input multiple-output (MIMO) systems, ultra-reliable low-latency communication (URLLC), physical layer design of extremely large aperture array (ELAA) systems and stochastic process.
\end{IEEEbiography}

\begin{IEEEbiography}[{\includegraphics[width=1in,height=1.25in,clip]{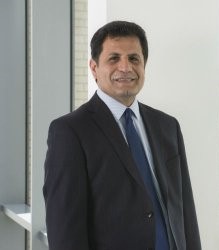}}]{Rahim Tafazolli}
	(Senior Member, IEEE) is Regius Professor of Electronic Engineering, Professor of Mobile and Satellite Communications, Founder and Director of 5GIC, 6GIC and ICS (Institute for Communication System) at the University of Surrey. He has over 30 years of experience in digital communications research and teaching. He has authored and co-authored more than 1000 research publications and is regularly invited to deliver keynote talks and distinguished lectures to international conferences and workshops. He was the leader of study on “grand challenges in IoT" (Internet of Things) in the UK, 2011-2012, for RCUK (Research Council UK) and the UK TSB (Technology Strategy Board). He is the Editor of two books on Technologies for Wireless Future (Wiley) vol. 1, in 2004 and vol. 2, in 2006. He holds Fellowship of Royal Academy of Engineering, Institute of Engineering and Technology (IET) as well as that of Wireless World Research Forum. He was also awarded the 28th KIA Laureate Award- 2015 for his contribution to communications technology.
\end{IEEEbiography}

\end{document}